%% file: main.tex
\documentclass[fleqn,usenatbib, trackchanges]{mnras}
\usepackage[T1]{fontenc}
\usepackage{lineno}

\DeclareRobustCommand{\VAN}[3]{#2}
\let\VANthebibliography\thebibliography
\def\thebibliography{\DeclareRobustCommand{\VAN}[3]{##3}\VANthebibliography}

\usepackage{graphicx}	
\usepackage{amsmath}	
\usepackage{amssymb}	
\usepackage{verbatim}
\usepackage{array}  
\usepackage{amstext} 
\usepackage{longtable}
\usepackage{supertabular}
\usepackage{soul} 
\usepackage[dvipsnames]{xcolor}
\usepackage{gensymb}
\usepackage[capitalize]{cleveref}
\usepackage{xcolor}

\newcommand{\thisStar}{2MASS J00512646-1053170}
\newcommand{\thisStarShort}{J0051-1053}
\newcommand{\rproc}{$r$-process}
\newcommand{\sproc}{$s$-process}
\newcommand{\eTeff}{$T_\mathrm{eff}$}
\newcommand{\logg}{$\log\ g$}
\newcommand{\vmicro}{$\xi$}
\newcommand{\loggf}{$\log\ gf$}
\newcommand{\eps}{$\log\epsilon$}
\newcommand{\Msun}{\mathrm{M}_\odot}

\newcolumntype{L}{>{$}l<{$}} 
\newcolumntype{C}{>{$}c<{$}}

\title[UV and Optical Spectroscopy of \thisStarShort]{The $R$-Process Alliance: Detailed Chemical Composition of an \textit{R}-Process Enhanced Star with UV and Optical Spectroscopy}

\author[Shah et al]{Shivani P. Shah,$^{1}$\thanks{E-mail: shivani.shah@ufl.edu}
Rana Ezzeddine,$^{1}$
Ian U. Roederer,$^{2,3,4}$
Terese T. Hansen,$^{5}$
Vinicius M. Placco,$^{6}$
\newauthor 
Timothy C. Beers,$^{7,4}$ 
Anna Frebel,$^{8,4}$
Alexander P. Ji,$^{9,10}$ 
Erika M. Holmbeck,$^{11}$ 
\newauthor
Jennifer Marshall,$^{12}$ 
Charli M. Sakari$^{13}$
\\
$^{1}$Department of Astronomy, University of Florida, 211 Bryant Space Science Center, Gainesville, FL 32601, USA\\
$^{2}$Department of Physics, North Carolina State University, Raleigh, NC 27695, USA\\
$^{3}$Department of Astronomy, University of Michigan,
Ann Arbor, MI 48109, USA\\
$^4$Joint Institute for Nuclear Astrophysics -- Center for the
Evolution of the Elements (JINA-CEE), USA\\
$^5$Department of Astronomy, Stockholm University, AlbaNova University Centre, SE-106 91 Stockholm, Sweden\\
$^6$NSF’s NOIRLab, Tucson, AZ 85719, USA\\
$^7$Department of Physics and Astronomy, University of Notre Dame, Notre Dame, IN 46556, USA\\
$^8$Department of Physics and Kavli Institute for Astrophysics and Space Research, Massachusetts Institute of Technology, Cambridge,\\ MA 02139, USA\\
$^9$Department of Astronomy \& Astrophysics, University of Chicago, 5640 S Ellis Avenue, Chicago IL 60637, USA\\
$^{10}$Kavli Institute for Cosmological Physics, University of Chicago, Chicago IL 60637, USA\\
$^{11}$The Observatories of the Carnegie Institution for Science, 813 Santa Barbara St, Pasadena, CA 91101, USA\\
$^{12}$Mitchell Institute for Fundamental Physics and Astronomy and Department of Physics and Astronomy, Texas A\&M University,\\ College Station, TX 77843-4242,
USA\\
$^{13}$Department of Physics \& Astronomy, San Francisco State University, San Francisco CA 94132, USA
}

\date{Accepted 2024 January 16. Received 2024 January 14; in original form 2023 December 19}

\pubyear{2024}

\begin{document}

\label{firstpage}
\pagerange{\pageref{firstpage}--\pageref{lastpage}}
\maketitle

\begin{abstract}
We present a detailed chemical-abundance analysis of a highly \rproc-enhanced (RPE) star, \thisStar, using high-resolution spectroscopic observations with \textit{Hubble Space Telescope}/STIS in the UV and Magellan/MIKE in the optical. We determined abundances for 41 elements in total, including 23 \rproc\ elements and rarely probed species such as \ion{Al}{II}, \ion{Ge}{I}, \ion{Mo}{II}, \ion{Cd}{I}, \ion{Os}{II}, \ion{Pt}{I}, and \ion{Au}{I}. We find that [Ge/Fe] $= +0.10$, which is an unusually high Ge enhancement for such a metal-poor star and indicates contribution from a production mechanism decoupled from that of Fe. We also find that this star has the highest Cd abundance observed for a metal-poor star to date. We find that the dispersion in the Cd abundances of metal-poor stars can be explained by the correlation of \ion{Cd}{I} abundances with the stellar parameters of the stars, indicating the presence of NLTE effects. We also report that this star is now only the 6th star with Au abundance determined. This result, along with abundances of Pt and Os, uphold the case for the extension of the universal \rproc\ pattern to the third \rproc\ peak and to Au. This study adds to the sparse but growing number of RPE stars with extensive chemical-abundance inventories and highlights the need for not only more abundance determinations of these rarely probed species, but also advances in theoretical NLTE and astrophysical studies to reliably understand the origin of \rproc\ elements.

\end{abstract}



\section{Introduction}

The rapid neutron-capture process (\rproc) is thought to be responsible for synthesizing about half the isotopes of elements heavier than zinc (atomic number, Z > 30) observed in the Solar System (S.S.) \citep{Cameron1957, B2FH}, with the slow neutron-capture process (\sproc) responsible for synthesizing the other half. 
Easily detected \rproc\ elements such as europium (Eu) have also been observed outside the S.S., in various Milky Way stars, stellar streams, and dwarf galaxies \citep[e.g.,][and references therein]{Venn+2004_MWrproc, Ji2016_RetII_APJ, DelgadoMena_2017_MWrproc, Marshal2019_TucanaIII, hansen2021_IndusStream, Ji2022_Typhon}. However, the primary astrophysical site(s) of \rproc\ nucleosynthesis is still unresolved, contributing to a substantial gap in our understanding of Galactic chemical enrichment and evolution \citep{CowanReview2021,Astro2020DecadalSurvey}.
\par 
Theoretical studies investigating the properties of \rproc\ astrophysical sites have typically relied on the S.S. \rproc\ pattern to understand the range of \rproc\ elements synthesized, and the relative quantities in which they are synthesized \cite[e.g.,][]{Goriely2001_multieventrprocmodel, Schatz2002, Farouqu2010_highentropywind, Lippuner2017_HMNS, Siegel2019_collapsars, Curtis2023_BHNS}. However, the S.S. \rproc\ pattern is not directly measured. Instead, it is obtained as the residual of  the S.S.'s total chemical-abundance pattern after accounting for a theoretically derived \sproc\ pattern \cite[e.g.,][]{Arlandini1999, Sneden2008_isotopes, Prantzos2020_SS}. In turn, the \sproc\ pattern is obtained by calibrating stellar evolution and Galactic chemical-evolution models to the meteoritic abundance of \sproc-only isotopes, with \sproc\ abundances of all other isotopes theoretically inferred  \citep{Roederer22}. Moreover, the S.S. \rproc\ abundance pattern represents only a single \rproc\ template, which is also the result of Galactic chemical evolution over billions of years.
\par
On the other hand, \rproc-enhanced (RPE) stars serve as more direct and reliable probe of \rproc\ nucleosynthesis events, as well as provide the opportunity of obtaining many different \rproc\ templates.
RPE stars have \rproc\ elemental abundances in excess of twice the Fe abundance as compared to the Sun, [Eu/Fe] > +0.3\footnote{$\mathrm{[A/B]} = \log(N_A/N_B)_\mathrm{Star} - \log({N_A/N_B})_\mathrm{Solar}$, where $N$ is the number density of the element.}  \citep{Beers&Christlieb2005Rev, rpa4}. RPE stars are typically very metal poor, with [Fe/H] $\lesssim -2.0$, and as a result, they have preserved the pristine chemical fingerprints of very few (in some cases just one) progenitor \rproc\ nucleosynthesis events \cite[][and references therein]{Frebel18_rev}. Additionally, the heavy-element abundance patterns of the RPE stars originate almost purely from the \rproc, with minimal contributions from other processes, if any. Given all of this, RPE stars offer a unique view of \rproc\ nucleosynthesis in the early Universe.
\par
To fully leverage RPE stars, it is important to determine the abundances for a wide range of their \rproc\ elements. Reliable abundances for the rare-earth \rproc\ elements ($55 < \mathrm{Z} < 71$) can be obtained relatively easily via ground-based optical observations \citep[e.g.,][]{Sneden2009_rareEarths, Gull2021_helmiStream}. On the other hand, to reliably determine abundances for most of the other elements, including lighter elements at and between the first and second \rproc\ peaks ($30 < \mathrm{Z} < 55$) and heavier elements at and around the third \rproc\ peak ($71 \lesssim\ \mathrm{Z} < 84$), supplemental space-based ultraviolet (UV) observations are required \citep[e.g.,][]{SiqueiraMello2013_CS31082, Roederer22}. In fact, robust abundance determinations for some of these elements, such as germanium (Ge, Z = 32), selenium (Se, Z = 33), cadmium (Cd, Z = 48), tellurium (Te, Z = 52), platinum (Pt, Z = 78), and gold (Au, Z = 79), have been possible solely because of UV observations by the \textit{Hubble Space Telescope} (HST) \citep{Cowan1996_PtOsPd_UV, Sneden1998_PtOsPb_UV, Cowan2002, Sneden2003_CS22892,
Roederer2010, Roederer2012_Te,Roederer&Lawlor2012, Roederer2014_AsSe,  SiqueiraMello2013_CS31082}. 
\par
UV observations of RPE stars are especially desired to characterize the signatures of a larger inventory of \rproc\ elements. Signatures of the light \rproc\ elements accessible via optical spectra (e.g., Sr, Y, Zr) have indicated that the abundances of these elements deviate from the "universal" \rproc\ pattern observed for the rare-earth and third-peak elements \citep[][and references therein]{Sneden2000,Cowan2005_largeSample, Francois2007, SiqueiraMello2014_rI, Ji2016_RetII_APJ, CowanReview2021}. The origin of these deviations is still unknown, with different astrophysical sites, conditions, and processes being considered \citep[e.g.,][]{Chiappini2011, Hansen2012, Wanajo2013, Holmbeck2019, Roederer2022b, Roe2023_science}. However, the full extent of these deviations is still not even well-established for some elements, such as Se, Cd, and Te, due to their scarce abundances, which require high-resolution UV spectra \citep{Roederer2022b}. The abundances of these elements, especially at and around the second \rproc\ peak are also crucial in constraining the effects of nuclear physical processes like fission cycling on the \rproc\ abundances \citep{Eichler2016, Vassh2019}. Similarly, even though the  heavier \rproc\ elements exhibit a "universal" abundance pattern, elements such as Pt and Au at the third \rproc\ peak have been observed in $\sim 15$ and $\sim5$ RPE stars, respectively, questioning the extent of the \rproc\ universality or alternatively holding undiscovered clues to the still unknown origin of the universality (although see \citealt{Roe2023_science}). 
\par
UV observations also offer the opportunity to investigate NLTE effects for the neutral species of elements such as Mg, Al, Co, Ni, Mo, and Os. Specifically, they enable access to the dominant species (usually first ions) of these elements, while their minority species (neutral atoms) are accessible via optical observations \citep{Roederer2010, Peterson2011_MoII, Roederer2021_AlII, Roederer22}. A subsequent comparison between the abundances of the dominant and the minority species can promote an empirical assessment of the theoretically predicted NLTE effects for the minority species of the elements, and thereby an assessment of the NLTE models themselves. 
\par
However, there are only a few RPE stars that have been analyzed with space-based UV observations and ground-based optical observations. These include HD 222925 \citep{Roederer2018_HD222925, Roederer22}, CS 31082-001 \citep{Cayrel2001_cs31082, Hill2002_CS31082, Plez2004_CS311082, Barbuy2011_CS31082, SiqueiraMello2013_CS31082}, CS 22892-052 \citep{Sneden2003_CS22892}, HD 108317 \citep{Roederer2012_HST, Roederer2014_largesample, Roederer2014_AsSe}, BD +17 3248 \citep{Cowan2002, Roederer2010}, HD 160617 \citep{Roederer2012_HD160617, Peterson+2020}, HD 84937 \citep{Peterson+2020}, and HD 19445 \citep{Peterson+2020}. The scarcity of such studies partly arises from the need of an RPE star to be bright in the NUV (e.g., GALEX \textit{NUV} $<15$)  in order to achieve the necessary signal-to-noise ratio \citep[e.g.,][]{Roederer2022b}. Additionally, since the strength of the absorption lines are decreased in such NUV-bright stars due to their higher effective temperatures, the stars also have to be sufficiently \rproc\ enhanced to obtain abundances for a wide range of \rproc\ elements. Studies of these stars have typically resulted in the determination of $\sim25$ to $35$ \rproc\ elemental abundances for these stars. An exception to this case is HD 222925, for which abundances of a record 42 \rproc\ elements were determined. While these studies have already resulted in important theoretical implications for \rproc-nucleosynthesis \citep[e.g.,][]{Roederer2022b, Holmbeck2023}, further advances in the field are still limited by the small number of stars that are studied in this manner. 
\par
In this paper, we present a detailed chemical-abundance analysis of a highly \rproc-enhanced star, \thisStar\ (hereafter \thisStarShort), using UV observations with the \textit{Space Telescope Imaging Spectrograph} (STIS) on board HST and optical observations with \textit{the Magellan Inamori Kyocera Echelle} (MIKE) instrument at the Magellan II telescope. This star is unique from most other RPE stars studied with UV and optical spectroscopy, since it is possibly the warmest star in the sample ($\sim6400$ K) which is also highly \rproc\ enhanced ([Eu/Fe]$\ \sim +1.30$) and very low in metallicity ([Fe/H]$\ \sim -2.30$). 
\par
This paper is organized as follows: In Section \ref{sec:pedigree}, we provide a brief overview of the discovery and previous literature studies of \thisStarShort. In Section \ref{sec:DataReduction}, we describe the data collection and reduction. Stellar parameter determination is outlined in Section \ref{sec:stellarParam}, while the linelist and atomic data used are specified in Section \ref{sec:linelist}. We describe the abundance determination of all the elements in Section \ref{sec:chemAbund}. We detail the detection threshold method in Section \ref{sec:detLimit}, and the uncertainty analysis in Section \ref{sec:uncertainty}. We discuss the results in Section \ref{sec:Discussion}, and conclude in Section \ref{sec:conclusion}.

\section{Pedigree of \thisStarShort\ and Its Possible Helmi Stream Membership}\label{sec:pedigree}
\thisStarShort\ was first identified as a candidate metal-poor star by the Hamburg/ESO Survey, and then confirmed to have [Fe/H]$\ = -2.43$ through medium-resolution ($R \sim 2000$) spectroscopic followup with the 4 m Blanco telescope at the Cerro Tololo Inter-American Observatory \citep{Frebel2006_HES}. Subsequently, \thisStarShort\ was identified as a possible member of the Helmi Stream by \cite{Beers2017_HES}. \thisStarShort\ was then identified by \cite{RPA3} as a highly \rproc-enhanced star with [Eu/Fe]$ = 1.34$ through higher-resolution spectroscopy with Magellan/MIKE at the Las Campanas Observatory, Chile. Given the importance of its chemical properties as a possible Helmi stream member, \thisStarShort\ was studied in further detail by \cite{Gull2021_helmiStream}, who obtained abundances for 12 \rproc\ elements. 
\par
We note that a more recent study by \cite{Koppelman2019}, who identified $\sim$600 potential members of the Helmi stream using Gaia DR2 kinematic parameters of over 8 million stars, did not identify \thisStarShort\ as one of the members. A kinematic analysis of \thisStarShort\ by G. Limberg following the method in \cite{Limberg2021_HelmiAbundances} with updated Gaia DR3 parameters also indicated that \thisStarShort\ is unlikely to be associated with the Helmi stream (private communication). Given this uncertainty, we do not discuss the membership of \thisStarShort\ further.

\section{Data Acquisition and Reduction}\label{sec:DataReduction} 

\subsection{Optical Data}
We observed \thisStarShort\ on 2016 October 14 (MJD = 57675.15855) with Magellan/MIKE \citep{Bernstein2003_MIKE} at the Las Campanas Observatory for a total exposure time of 1200s, resulting in S/N of 150 per pixel at 4000 \AA. We used the $0\farcs7$ slit with $2 \times 2$ binning, which yielded a measured resolving power of $R$ $\sim$28,400/26,800 in the blue and red arms, respectively. The blue- and red-arm spectra together cover a wavelength range of $3350$-$9500$ \,\AA. We reduced the spectra of the star using \texttt{CarPy} \citep{Kelson2000_CarPy1stref, Kelson2003_CarPy2ndref}, and corrected the radial velocity by cross-correlating against a rest-frame Magellan/MIKE spectrum of G 64-12. We determined a resulting heliocentric radial velocity of 56.91 km/s. To normalize the orders, we used \texttt{SMHr}\footnote{\url{https://github.com/eholmbeck/smhr-rpa/tree/py38-mpl313}}, specifically using a natural cubic spline function with sigma clipping and strong lines masked, which was followed by stitching the orders together to furnish the final spectrum.

\subsection{Near-Ultraviolet Data}
\thisStarShort\ was observed with HST/STIS \citep{STIS_citation1, STIS_citation2} on 2020 January 21, 22, and 23 \citep[Proposal ID: 15951]{hst2019_prop}. The star was observed with the E230M \'echelle grating centered at $\lambda2707$ \AA, providing a wavelength coverage from  $\sim2275-3119$ \AA, and with the $0\farcs2 \times 0\farcs2$ slit, providing $R\sim30,000$\footnote{\url{https://hst-docs.stsci.edu/stisihb/chapter-13-spectroscopic-reference-material/13-3-gratings/echelle-grating-e230m}}. The observations were made over 12 orbits (i.e., 12 continuous exposures), with 3 orbits in each visit. The total exposure time over the 12 orbits was $\sim$5.65h, resulting in a S/N of 65 at 2707 \AA. The spectra were automatically reduced by the \texttt{CALSTIS} software package, and we downloaded the processed spectra from the Mikulski Archive for Space Telescopes. We corrected for the radial velocity of each exposure by cross-calibrating a synthetic spectrum of the star generated with \texttt{MOOG} \citep{Sneden1973_MOOG}. We normalized the orders using a natural cubic spline function with sigma clipping, and co-added the normalized orders of all the exposures before stitching to furnish the final spectrum\footnote{\url{https://github.com/alexji/alexmods/blob/master/alexmods/specutils/continuum.py}}. 

\section{Stellar Parameters}\label{sec:stellarParam}
We determined the effective temperature (\eTeff) and surface gravity (\logg) of \thisStarShort\ photometrically, based on methods described in \cite{Roederer2018_HD222925} and \cite{Placco2020_J183}. We preferred to use photometric determinations of \eTeff\ and \logg\ since spectroscopic determinations based on LTE have been known to be inaccurate and requiring additional corrections \citep{Thevenin1999, Frebel2013_Teff, Ezzeddine2017, RPA3}. Moreover, these methods follow the $R$-Process Alliance convention for homogeneity. We briefly describe the methods used below. 
\par
We determined \eTeff\ using the color--[Fe/H]--\eTeff\ photometric relations of \citet{Casagrande+2010}, which require an estimate of the metallicity. We initially used [Fe/H] $= -1.97$, obtained with spectroscopic determination of the stellar parameters. With new \eTeff\ and \logg\ estimates obtained photometrically, we re-determined [Fe/H] using equivalent-width (EW) measurements of \ion{Fe}{I} lines in the optical. We repeated the \eTeff\ and \logg\ calculation using the photometric relations with [Fe/H] $\ = -2.33$. 
\par
We calculated \eTeff\ from the dereddened $V-J$, $V-H$, $V-K$, and $J-K$ colors. We used the $J, H,$ and $K$ magnitudes from 2MASS \citep[][\href{https://vizier.u-strasbg.fr/viz-bin/VizieR-3?-source=II/246}{{\texttt{Vizier catalog II/246}}}]{Cutri+2003_2MASSVizier} and the Johnson $V$ magnitude from DR9 of APASS \citep[][\href{https://vizier.u-strasbg.fr/viz-bin/VizieR-3?-source=II/336}{{\texttt{Vizier catalog II/336}}}]{HendenMunari_2004APASS}. We adopted the reddening value, $E(B-V)$, of 0.001 from \cite{Schlafly2011} for the line of sight of the star, with  the $A_{\lambda}$ extinction coefficient for the color bands from \cite{McCall2004}. We chose to not use the $B-V$ color, because the $B$-band is sensitive to the CH $G$-band in carbon-enhanced metal-poor stars, which was not taken into account in the photometric relations. However, as shown below, \thisStarShort\ was not found to be carbon enhanced. As described in \citet{Roederer2018_HD222925}, we then calculated \eTeff\ for each color band by drawing the input parameters (magnitudes, reddening, and metallicity) $10^4$ times from their corresponding error distributions, which we assumed to be Gaussian. We used the median value of the resulting \eTeff\ distribution. For the final \eTeff, we used the weighted average of \eTeff\ from all the color bands. For the total uncertainty on \eTeff, we used the uncertainty of the weighted average. As a result, we obtained \eTeff$= 6440\ \pm\ 82$~K. 
\par
We calculated \logg\ using the following fundamental relation: 
\begin{align}
    \log\ g  & = 4 \log T_\mathrm{eff} + \log(M/\Msun) - 10.61 + 0.4\cdot(BC_{V})\nonumber\\
           & +\ V - 5\log (d) + 5 - 3.1 \cdot E(B-V) - M_{\rm bol, \odot} 
\end{align}
For $M$, the mass of the star, we assumed $0.8 \pm 0.08$ $\Msun$. For BC$_V$, the bolometric correction in the $V$-band, we used $-0.22$ \citep{Casagrande&VandenBerg2014_BCV}. We obtained the distance $d = 3.11$ pc from \cite{Bailer-Jones2021}.
M$_{\rm bol,\odot}$ is the Solar bolometric magnitude, equal to $4.75$. We calculated the constant $10.61$ from the Solar constants log(\eTeff)$_{\odot} = 3.7617$ and \logg$_{\odot} = 4.438$. We estimated \logg\ by drawing these input parameters 10$^4$ times from their error distribution and taking the median of the resulting distribution of \logg. For the uncertainty on \logg, we used the standard deviation of the distribution. We note that, in order to take the error distribution of \eTeff\ into account as one of the inputs to the \logg\ calculation, we added $150$~K in quadrature with the uncertainty calculated above ($82$~K) to account for other possible systematic uncertainties. This choice has a minimal impact on the \logg\ value. Finally, we obtained \logg$\ = 4.02\ \pm\ 0.07$ dex.
\par
For determining \vmicro\ and [Fe/H], we used the EWs of \ion{Fe}{I} and \ion{Fe}{II} lines, along with the \eTeff\ and \logg\ values obtained above. For this, we used \texttt{SMHr}\footnote{\url{https://github.com/andycasey/smhr}}, the next-generation spectroscopic analysis tool of \texttt{SMH} \citep{Casey2014_SMH}. \texttt{SMHr} uses the radiative transfer code \texttt{MOOG} \citep{Sneden1973_MOOG}, which we used with the proper treatment of scattering included\footnote{\url{https://github.com/alexji/moog17scat}} \citep{Sobeck2011_MOOG}. For the stellar model atmosphere, we employed the ATLAS9 1D plane parallel LTE grid \citep{Castelli&Kurucz2004_ATLAS9} in \texttt{SMHr}. To determine \vmicro, we minimized the trend in the abundances of the optical \ion{Fe}{I} lines and their reduced EWs. We obtained \vmicro$\ =\ 1.59$ km/s. We assumed a fiducial uncertainty of 0.2 km/s on \vmicro. For [Fe/H], we report the resulting mean abundance of the UV and optical \ion{Fe}{II} lines, with \eTeff, \logg, and \vmicro\ fixed to the above values, which is [Fe/H]$\ = -2.34$. We assumed a fiducial uncertainty of $0.20$ dex on [Fe/H] as well.
\par
Given the \eTeff\ and \logg\ of \thisStarShort, and its position on the Hertzprung-Russell diagram, we identify the star to be near the end of the turn-off phase and at the beginning of the subgiant phase. For the purpose of this paper, we refer to it as a turn-off star.

\section{Linelist and Atomic Data}\label{sec:linelist}
We used \texttt{linemake}\footnote{\url{https://github.com/vmplacco/linemake/tree/ch_masseron}} \citep{linemake_2021_code, linemake2021} to generate the linelists for the UV and optical absorption lines. We specifically used the updated parameters of the CH transitions from \cite{Masseron2014_fixed}. The linelist included hyperfine splitting structure (HFS) for relevant transitions. Following \cite{Roederer22}, we included the \ion{Au}{I} line at 2376.28 \AA, with oscillator strength (\loggf-value) from \cite{Zhang+2018_AuI}. We also updated the \loggf-values of \ion{Hf}{II} lines and included additional \ion{Hf}{II} lines from \cite{DenHartog_2021_HfII}.
\par
For abundance determination, we investigated absorption lines used by \cite{Placco2015_HST}, \cite{Roederer2018_HD222925}, \cite{Ji2020_s5}, and \cite{Roederer22}. We also investigated additional absorption lines of light and heavy elements in the UV from the \textit{National Institute of Standard and Technology Atomic Spectra Database} (NIST ASD) \citep{NIST_ASD}. We present the final list of absorption lines used for abundance determination in Table \ref{tab:all_lines}, along with the corresponding atomic parameters of the lines. We used a total of 113 transitions in the UV and 402 transitions in the optical.

\section{Chemical Abundances}\label{sec:chemAbund}
As described in Section \ref{sec:linelist}, we used \texttt{SMHr}, along with \texttt{MOOG} and ATLAS9 stellar-atmosphere grid, for line-by-line spectral analysis and abundance determination. We determined abundances using EW measurements for \ion{Na}{I}, \ion{K}{I}, \ion{Ca}{I}, \ion{Ti}{II}, \ion{Fe}{I}, \ion{Fe}{II}, and \ion{Ni}{I} lines in the optical. For all the other lines, we used spectral synthesis. We used the isotopic ratios of the \rproc\ elements from \cite{Sneden2008_isotopes}. 
\par
For an overview, we present various results of the abundance analysis in tables. Table \ref{tab:all_lines} lists, for all the absorption lines used, the atomic parameters (wavelength, excitation potential, and \loggf-value), the abundance determination technique used (EW or spectral synthesis), the resulting abundance, and the systematic uncertainty on the abundance. Table \ref{tab:UVOP_abund} lists the mean abundances of each species from the UV and optical spectra separately, along with the statistical uncertainty. Table \ref{tab:adopted_abund} lists the adopted abundance for each element, along with the resulting [X/Fe] abundance ratio, the statistical uncertainty, the systematic uncertainty, and the total uncertainty. 
\par
In general, we adopted the abundance of the ionized species, if available. If unavailable, we adopted the NLTE-corrected abundance of the neutral species, if the NLTE correction grids were available in the literature. If neither of these cases applied, we adopted the LTE abundance of the neutral species. If available, we generally combined the UV and optical lines for abundance determination of each species by taking the average of all the lines. We further detail the abundance determination for all the elements below. 

\input{tables/abund_latex_wUnc.txt}

\subsection{Li}
We detected a strong \ion{Li}{I} absorption feature at $6707.80$ \AA. We did not detect Li absorption at any other \ion{Li}{I} transition lines \citep[e.g.,][]{Kowkabany2022_li}. We used spectral synthesis to fit the $\lambda6707$ line and determined \eps(Li)$\ = 2.37$. This absorption feature has a potential contribution from both the $^{6}$Li and $^{7}$Li isotopes. We used an isotopic ratio of $^{6}$Li/$^{7}$Li$ = 0.00$, since any reasonable isotopic contribution from $^6$Li is undetectable at the resolution of our spectrum. For instance, increasing the isotopic ratio to $0.01$, which is the upper limit suggested by \cite{Prantzos2012_Li} for metal-poor stars based on Galactic chemical evolution, did not change the line profile in any detectable manner. Moreover, it is recommended to implement 3D and NLTE models to reliably constrain the isotopic ratio from the spectral signature \citep{Lind2013_Li}, which is beyond the scope of this work. However, we obtain a NLTE correction of $-0.018$ using \texttt{Breidablik}\footnote{\url{https://github.com/ellawang44/Breidablik}} \citep{Wang2021_Breidablik}, which is an interpolation routine to estimate the NLTE correction of the Li abundance based on the EW of the line. For the final Li abundance, we adopted the NLTE-corrected abundance of \eps(Li)$\ = 2.35$. 

\input{tables/adopted_abund_latex.txt}

\subsection{C and N }
We determined \eps(C)$\ = 6.70$ using the CH $G$-band at $4313$ \AA. We also obtained an identical abundance using the \ion{C}{I} $\lambda2964.85$ line. However, we don't find this line reliable in our spectrum since it is not very well resolved. Moreover, as pointed out by \cite{Roederer22}, the accuracy of the line's \loggf\ value is graded a D (<50\%; $\pm0.30$ dex) by NIST. Therefore, we adopted the C abundance from the CH $G$-band only. We further considered a correction to the C abundance due to possible CN processing as described in \cite{Placco2014_C}, but find a correction of 0.0 dex\footnote{Using the online tool available at \url{http://vplacco.pythonanywhere.com/}} due to the relatively early evolutionary stage of the star. 
\par
For N, we used the $\lambda3876$ CN molecular band. However, the CN absorption features were very weak, so we determined a $3\sigma$ upper limit, obtaining \eps(N)$\ < 7.73$.

\subsection{Na, Al, and K}
We determined \eps(Na)$\ = 3.99$ using EW measurements of four \ion{Na}{I} lines including the $\lambda5889/\lambda5895$ and $\lambda8183/\lambda8194$ doublets. We checked for telluric blending with the $\lambda$8183/$\lambda$8194 doublet using the telluric spectrum provided with the Arcturus Atlas by \cite{ArcturusAlas_telluric} and did not identify any strong blending. Moreover, the four \ion{Na}{I} sodium lines render abundances in excellent agreement. We also estimated a NLTE correction of $-0.10$ dex, based on NLTE correction girds provided by \cite{Lind2022_NaMgAlNLE}. For the final Na abundance, we adopted the NLTE-corrected abundance of \eps(Na)$\ = 4.09$. 
\par
\begin{figure}
    \centering
    \includegraphics[width = 0.46\textwidth]{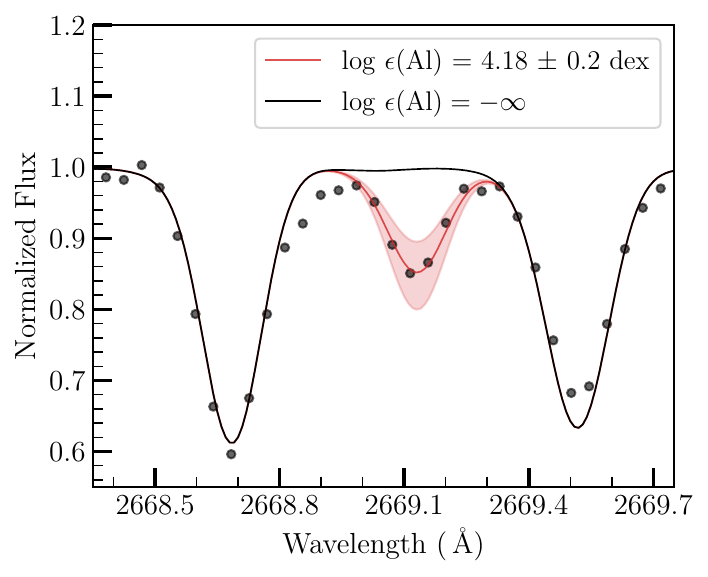}
    \caption{The best-fit spectral synthesis model for the \ion{Al}{II} line at 2669.16 \AA\ is shown with a red-solid line, with the observed data shown in black points. The red-shaded region depicts abundance variation within $\pm0.2$ dex of the best-fit abundance. The black-solid line traces the synthetic model with no contribution from Al.}
    \label{fig:Al}
\end{figure}

For Al, we could determine the abundance for two \ion{Al}{I} lines in the optical and one \ion{Al}{II} line in the UV using spectral synthesis. For the mean \ion{Al}{I} abundance, we obtained \eps(Al)$= 3.32$. Specifically, we used spectral synthesis of \ion{Al}{I} lines at $3944.00$ \AA\,\ and $3961.52$ \AA. On the other hand, we determined a much higher abundance of \eps(Al)$\ = 4.21$ using the best-fit spectral synthesis model for the $\lambda2669.16$ \ion{Al}{II} line, as shown in Figure \ref{fig:Al}\citep{Roederer2021_AlII}. The reason for this discrepancy has been proposed to be NLTE effects on the abundances of low-excitation \ion{Al}{I} lines, such as those used here, on the order of  $\sim+0.4$ dex for metal-poor main-sequence turn-off stars \citep{Nordlander2017_AlNLTE, Roederer2021_AlII}. On the other hand, the ground state $\lambda2669.16$ \ion{Al}{II} line is considered to be forming in LTE, providing a more faithful Al abundance determination \citep{Mashonkina2016_AlSiNLTE, Roederer2021_AlII}. Indeed, we estimated a NLTE correction of $+0.47$ dex for \ion{Al}{I} based on the NLTE correction grids provided by \cite{Mashonkina2016_AlSiNLTE}. However, in spite of this large NLTE correction, the NLTE corrected \ion{Al}{I} is still much lower than the \ion{Al}{II} abundance. Therefore, for the final Al abundance, we adopted the \ion{Al}{II} abundance of \eps(Al)$\ = 4.21$. 
\par
We determined \eps(K)$= 3.33$ using EW measurement of the \ion{K}{I} line at $7698.96$ \AA. While we also detected absorption at the $\lambda7664.90$ \ion{K}{I} line, it appears to be contaminated with telluric blends. We estimated a NLTE correction of $-0.28$ dex for the \ion{K}{I} abundance, based on the NLTE grids provided by \cite{Takeda2002_KI} for the $\lambda7698$ \ion{K}{I} line. For the final K abundance, we adopted the NLTE-corrected abundance of \ion{K}{I}, which was \eps(K)$\ = 3.05$.

\subsection{$\alpha$-Elements: O, Mg, Si, S, and Ca}
For O, we determined \eps(O)$= 7.08$ using spectral synthesis of the \ion{O}{I} triplet near $7770$ \AA. We also estimated a NLTE correction of $-0.10$ dex, based on the 1D NLTE correction grids provided by \cite{Amarsi2016_ONLTE} for the middle \ion{O}{I} triplet line in a turn-off star. For the final O abundance, we adopted the NLTE-corrected abundance of \eps(O)$ = 6.08$. 
\par
For Mg, we determined the abundances for 7 \ion{Mg}{I} lines in the optical, 2 \ion{Mg}{I} lines in the UV, and  one \ion{Mg}{II} line in the UV, all using spectral synthesis. We obtained \eps(Mg)$ = 5.73$ and \eps(Mg)$ = 5.72$, with the optical and UV \ion{Mg}{I} lines, respectively. Furthermore, we corrected the mean \ion{Mg}{I} abundance from these 9 lines for NLTE effects using the \ion{Mg}{I} grid provided by \cite{Lind2022_NaMgAlNLE}. We obtained a correction of +0.10 dex, bringing the NLTE \ion{Mg}{I} abundance to \eps(Mg)$ = 5.83$. For the \ion{Mg}{II} line at $\lambda2828$, we obtained \eps(Mg)$= 5.70$, in agreement with the LTE as well as the NLTE abundance \ion{Mg}{I} abundance. For the final Mg abundance, we adopted the \ion{Mg}{II} abundance. 
\par
For Si, we determined abundances for \ion{Si}{I} and \ion{Si}{II} lines in the UV and optical spectra using spectral synthesis. While we detected several \ion{Si}{I} lines in the UV spectrum, most were too strong for reliable abundance determination. Aside from the strong lines, two \ion{Si}{I} lines at $2438.77$ \AA\ and $2443.36$ \AA\ yielded Si abundance $\sim0.4$ dex higher than that determined from other \ion{Si}{I} and \ion{Si}{II} lines. It is unclear why this is the case, since the excitation potential of these lines, 0.00 and 0.01 eV, respectively, are only slightly lower than that of the other \ion{Si}{I} lines investigated, which have excitation potentials in the range of 0.78 - 1.91 eV. Therefore, we excluded these lines for the purpose of abundance determination, and suggest caution when using these lines. As a result, we used one line from the UV and one from the optical to obtain a mean \ion{Si}{I} abundance of \eps(Si)$ = 5.56$. We also estimated a NLTE correction for the optical \ion{Si}{I} line at $3905.53$ \AA\ using the MPIA-based NLTE correction tool\footnote{\url{https://nlte.mpia.de/gui-siuAC_secE.php}}, which yielded a correction of $+0.003$ dex based on the NLTE model grids provided by \cite{Bergemann2013_SiNLTE}. We assume a similar NLTE correction for the UV \ion{Si}{I} line and consider $+0.003$ dex to be the NLTE correction for the mean \ion{Si}{I} LTE abundance.  
\par
For the mean \ion{Si}{II} abundance, we used two \ion{Si}{II} lines in the UV and one \ion{Si}{II} line in the optical, which yielded \eps(Si)$ = 5.55$. We find that the \ion{Si}{II} abundance is in excellent agreement with the \ion{Si}{I} LTE abundance. On the other hand, \cite{Roe2016_BD44493} and \cite{Roederer22} found that the low-excitation \ion{Si}{I} lines yielded lower Si abundances than the high-excitation \ion{Si}{I} lines and the \ion{Si}{II} lines. We suspect that the better agreement observed for \thisStarShort\ might be in part due to the higher \logg\ of the star, resulting in lower NLTE effects for the low-excitation \ion{Si}{I} lines. For the final Si abundance, we adopted the \ion{Si}{II} abundance. 
\par
We inspected three \ion{S}{I} lines in the $\lambda6700$ region \citep{Roederer22}, but could not determine a reliable S abundance or upper limit.
\par
We determined the Ca abundance using EW measurements of 24 \ion{Ca}{I} lines in the optical and obtained \eps(Ca)$ = 4.48$. We also included a NLTE correction of $+0.26$ dex, as computed by \cite{Mashonkina2017_CaIandII_NLTE} for HD 84937, which has similar stellar parameters as \thisStarShort. For the final Ca abundance, we used the NLTE-corrected value of \eps(Ca)$ = 4.74$. 
\par

\subsection{Fe-group Elements: Sc, Ti, V, Cr, Mn, Fe, Co, Ni, and Zn}
We determined the Sc abundance using spectral synthesis of 12 \ion{Sc}{II} lines in the optical and one \ion{Sc}{II} line in the UV, which yielded \eps(Sc)$ = 1.10$ and \eps(Sc)$ = 1.13$, respectively. We used all 13 lines to determine the mean Sc abundance of \thisStarShort.
\par
We determined the Ti abundance using EW measurements of 34 \ion{Ti}{II} lines in the optical, which yielded \eps(Ti)$= 3.12$, and spectral synthesis of 5 \ion{Ti}{II} lines in the UV, which yielded \eps(Ti)$= 3.10$. We also determined a similar abundance of \eps(Ti)$ = 3.20$ from EW measurements of 12 \ion{Ti}{I} lines in the optical. Based on the NLTE analysis of Ti lines in HD 84937 by \cite{Sitnova2020_TiIandII_NLTE}, we estimated a NLTE correction of $+0.14$ and $+0.04$ dex for \ion{Ti}{I} and \ion{Ti}{II} in \thisStarShort, respectively. Given the negligible NLTE effects on \ion{Ti}{II}, for the final Ti abundance we adopted the LTE \ion{Ti}{II} abundance of \eps(Ti)$ = 3.12$. 
\par
We determined the V abundance using spectral synthesis of 7 \ion{V}{II} lines in the optical and 6 \ion{V}{II} lines in UV, which yielded \eps(V)$= 1.97$ and \eps(V)$= 1.99$, respectively. We were also able to determine \eps(V)$= 2.05$ using the $\lambda4111$ \ion{V}{I} line, which agrees very well with the \ion{V}{II} abundance. We note that NLTE grids for \ion{V}{I} and \ion{V}{II} are presently not available in the literature. For the final V abundance, we adopted the mean abundance from the 13 \ion{V}{II} lines.
\par
For Cr, we determined abundances for \ion{Cr}{I} and \ion{Cr}{II} lines in the UV and optical spectra with spectral synthesis. We determined \eps(Cr)$= 3.31$ with \ion{Cr}{I} lines and \eps(Cr)$= 3.42$ using the \ion{Cr}{II} line, which agree. We note that the mean \ion{Cr}{II} abundance from 11 lines in the UV is $+0.15$ dex higher than the mean \ion{Cr}{II} abundance from 3 lines in the optical. However, since the discrepancy is on the order of uncertainties on the Cr abundance ($\sim0.10$ dex), we don't suspect anything unusual at play. We also considered a NLTE correction of $+0.25$ dex and $+0.04$ dex for \ion{Cr}{I} and \ion{Cr}{II}, respectively, based on the NLTE analysis of HD 84937 by \cite{Bergemann_2010Cr}. Since the NLTE correction for \ion{Cr}{II} is negligible, for the final Cr abundance we adopted the mean LTE \ion{Cr}{II} abundance from 11 \ion{Cr}{II} lines in UV and 3 \ion{Cr}{II} lines in optical, which was \eps(Cr)$\ = 3.43$.
\par
For Mn, we determined abundances for \ion{Mn}{I} and \ion{Mn}{II} lines in the UV and optical spectra. We avoided the triplet \ion{Mn}{I} resonance lines in the $\lambda4030$ region since these yielded abundances $\sim0.20$ dex lower than the higher-excitation lines. The systematically lower abundances obtained with the Mn triplet, relative to other higher-excitation \ion{Mn}{I} lines, has been well-documented \citep[e.g.,][]{Cayrel2004, Roederer2010_Mn, Sneden2023_FePeak}. We find that the mean \ion{Mn}{II} abundance is slightly higher than the mean \ion{Mn}{I} abundance, but the discrepancy is within uncertainties. We estimated a NLTE correction of $+0.28$ and $-0.03$ dex for the \ion{Mn}{I} and \ion{Mn}{II} abundances, respectively, based on the NLTE grids provided by \cite{Bergemann2019_Mn3DNLTE}. For the final Mn abundance, we adopted the mean LTE \ion{Mn}{II} abundance  of \eps(Mn)$\ = 2.91$ from 2 UV lines and 4 optical lines.
\par
For Fe, we determined abundance for \ion{Fe}{I} and \ion{Fe}{II} lines in the UV using spectral synthesis and for \ion{Fe}{I} and \ion{Fe}{II} lines in the optical using EW measurements. For \ion{Fe}{I}, we determined abundances for 21 lines in the UV and 156 lines in the optical, and obtained a mean \ion{Fe}{I} abundance of \eps(Fe)$ = 5.19$. For \ion{Fe}{II}, we determined abundances for 22 lines in the UV and 14 lines in the optical, and obtained a mean \ion{Fe}{II} abundance of \eps(Fe)$ = 5.15$, which agrees very well with the mean \ion{Fe}{I} abundance. We note that the mean \ion{Fe}{II} abundance with the UV lines is $+0.13$ higher than the mean \ion{Fe}{II} abundance obtained with the optical lines (see Table \ref{tab:UVOP_abund}). However, since the standard deviation in the abundances of the UV lines is $0.12$, on the same order as the difference in the mean UV and optical abundances, we don't consider this discrepancy to be of concern. We also considered a NLTE correction of $+0.17$ dex for \ion{Fe}{I} and $0.0$ dex for \ion{Fe}{II}, based on the NLTE analysis of HD 84937 by \cite{Amarsi2016_FeNLTE}. For the final Fe abundance, we adopted the LTE \ion{Fe}{II} abundance from the UV and optical lines.  
\par
We determined abundances for \ion{Co}{I} lines in the UV and optical spectra as well as for \ion{Co}{II} lines, which are available only in the UV spectrum. We obtained a significantly  higher mean \ion{Co}{I} abundance of \eps(Co)$ = 2.88$ than the mean \ion{Co}{II} abundance of \eps(Co)$ = 2.61$. A similar discrepancy was observed by \cite{Cowan2020_FePeakElements} for three other turn-off stars with HST/STIS UV spectra.  On the other hand, \cite{Roederer22} did not observe such a discrepancy in the giant HD 222925. We further obtained a NLTE correction on the order of $+0.89$ dex for the \ion{Co}{I} lines used here using the \cite{Bergemann2010_Co} grids\footnote{We used the MPIA NLTE tool online at \url{https://nlte.mpia.de/gui-siuAC_secE.php} for interpolating the grids}, which exacerbates the discrepancy. We note that given the high NLTE corrections obtained and the absence of an explicit analysis of a turn-off metal-poor star in the study, it is unclear whether the grid reliably extends to temperatures as high as $6400$ K and metallicity as low as $-2.28$. We are expecting new NLTE studies for Co in the near future. Nevertheless, we adopted the mean abundance from 9 \ion{Co}{II} lines in the UV. 
\par
We determined abundances for \ion{Ni}{I} lines in the UV with spectral synthesis and for \ion{Ni}{I} lines in the optical with EW measurements. We also determined abundances for \ion{Ni}{II} lines in the UV using spectral synthesis. The mean \ion{Ni}{I} abundance of \eps(Ni)$= 3.94$ from 18 lines in the optical and 3 lines in the UV is the same as the \ion{Ni}{II} abundance of from 18 lines in the UV, indicating small NLTE effects for \ion{Ni}{I}. For the final Ni abundance, we adopted the mean \ion{Ni}{II} abundance of \eps(Ni)$ = 3.94$. 
\par
We looked for a \ion{Cu}{I} signature at the $\lambda5105$ line, however, we could not determine a reliable detection. 
\par
Lastly, for this group of elements, we determined \eps(Zn)$= 2.32$ with spectral synthesis of two \ion{Zn}{I} lines in the optical. We estimated a $+0.18$ dex NLTE correction on the Zn abundance using the NLTE grids provided by \cite{Sitnova2022_ZnNLTE} for the $\lambda4810$ \ion{Zn}{I} line. For the final Zn abundance, we adopted the NLTE-corrected abundance of \ion{Zn}{I}, which results in \eps(Zn)$ = 2.50$. 
\begin{figure*}
    \centering
    \includegraphics[width =\textwidth]{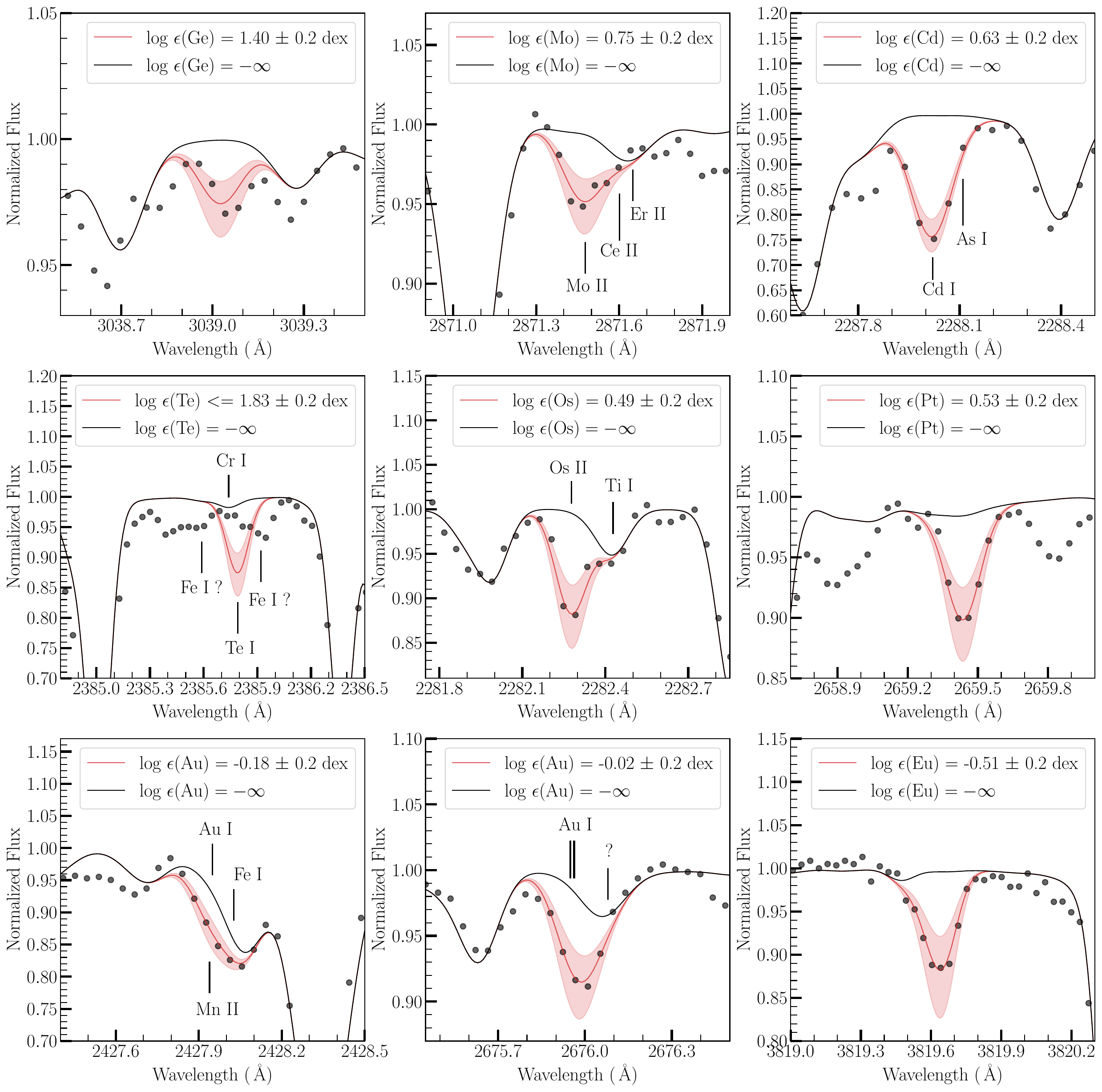}
    \caption{Spectral synthesis fits to absorption lines of various \rproc\ elements. The red-solid line traces the best-fit synthetic model to the observed data in black points. The red-shaded shaded region depicts abundance variation within $\pm0.2$ dex of the best-fit abundance. The black-solid line traces the synthetic model with no contribution from the relevant element. Important neighboring absorption lines are also labeled.}
    \label{fig:synFits}
\end{figure*}   

\subsection{Elements at the First \textit{R}-Process Peak}
Of the elements near the first \textit{r}-process peak, including Ga, Ge, As, and Se, the transitions of Ga and Se were out of the spectral range. For Ge, we determined \eps(Ge)$\ = 1.40$ using the $\lambda3039$ \ion{Ge}{I} line. We show the spectral synthesis fit to the region in Figure \ref{fig:synFits}. The spectral synthesis fit is not exact, possibly due to noise in the spectrum. Therefore, we assigned an additional $0.10$ dex uncertainty for the Ge abundance. For As, we could only determine a $3\sigma$ upper limit of \eps(As)$< 0.84$ using the $\lambda2288$ \ion{As}{I} line.

\begin{figure*}
	\includegraphics[width = 0.7\textwidth]{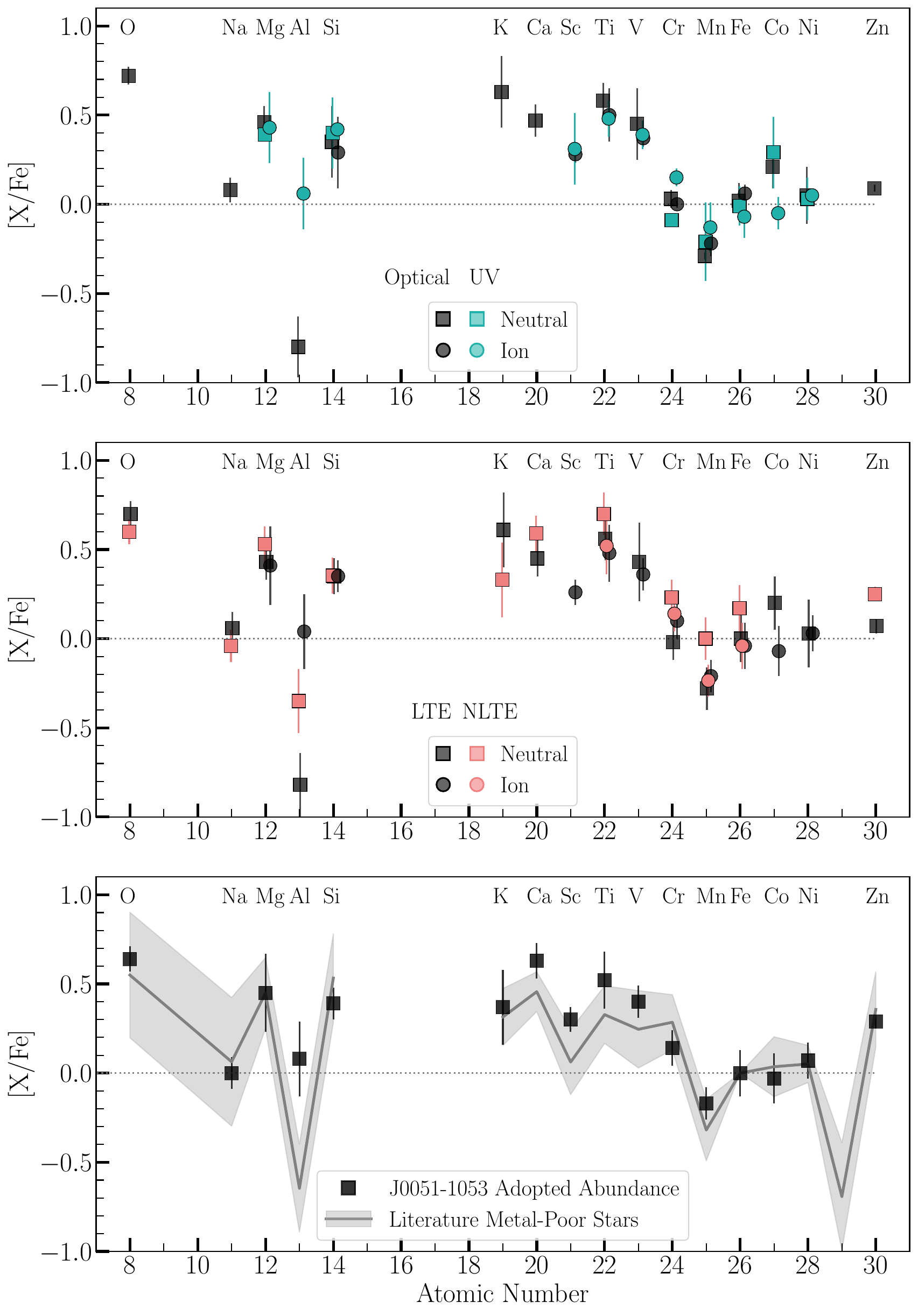}
    \caption{Top Panel: Mean [X/Fe] abundances of light-element species from the UV (green points) and optical spectra (black points). Middle Panel: Mean LTE (black points) and NLTE-corrected (red points) abundances of light-element species. Bottom Panel: The adopted abundances of the light elements for \thisStarShort\ are shown with black-square points. The black-solid line traces the mean [X/Fe] abundances of these elements for the metal-poor stars analyzed in \protect\cite{Roederer2014_largesample}. The grey-shaded region traces the corresponding standard deviation in the [X/Fe] abundances for the \protect \cite{Roederer2014_largesample} sample.}
    \label{fig:Fe_groupElements}
\end{figure*}

\subsection{Elements Between the First and Second \textit{R}-Process Peaks}\label{sec:betweenpeaks}
These elements include Rb (Z = 37), Sr (Z = 38), Y (Z = 39), and Zr (Z = 40). For Rb, we could not detect significant absorption, and thereby determined a 3$\sigma$ upper limit using the $\lambda7947$ \ion{Rb}{I} line. For Sr, we determined \eps(Sr)$\ = 1.05$ using two \ion{Sr}{II} transitions in the optical. For Y, we obtained \eps(Y)$\ = 0.58$ using one \ion{Y}{II} line in the UV and \eps(Y)$\ = 0.46$ using 9 \ion{Y}{II} lines in the optical. Since the mean UV and the mean optical \ion{Y}{II} abundances agree within their statistical uncertainties, we adopted the mean from all 10 lines as the Y abundance. Similarly, we determined the Zr abundance for 5 \ion{Zr}{II} lines in the UV (\eps(Zr)$= 1.09$) and 10 \ion{Zr}{II} lines in the optical (\eps(Zr)$= 1.15$), and we adopted the mean of all 15 lines as the Zr abundance. 
\par
For Nb, we detected only a faint absorption signature for the \ion{Nb}{II} at $2927.81$ \AA\  in the UV. Moreover, the blends were neither resolved nor constrained well by our synthetic spectrum. As a result, we could only determine a 3$\sigma$  upper limit. For Mo, we determined \eps(Mo)$\ = 0.71$ using the \ion{Mo}{II} line at $2871.51$ \AA\ in the UV. The other Mo lines were too weak for abundance determination, including the \ion{Mo}{I} transitions in the optical. Figure \ref{fig:synFits} shows the spectral synthesis fit to the $\lambda2971$ line. 
\par
We did not detect any clean strong absorption line of \ion{Ru}{I} or \ion{Ru}{II} in the UV or in the optical. Therefore, we determined a 3$\sigma$ upper limit for the Ru abundance using the $\lambda2456$ \ion{Ru}{II} line. Similarly for Rh, we investigated 6 \ion{Rh}{I} lines in the optical, but did not detect a reliable signature for abundance determination. Therefore, we determined a 3$\sigma$ upper limit for Rh using the $\lambda3435$ line. The trend continued for Pd and Ag, for which we could also only determine 3$\sigma$ upper limits using the $\lambda3405$ \ion{Pd}{I} and $\lambda3382$ \ion{Ag}{I} lines. In particular, for the \ion{Pd}{I} and \ion{Ag}{I} lines, the neighboring blends were not well-constrained or resolved.
\par
For Cd, we determined \eps(Cd)$\ = 0.63$ with the $\lambda2288$ \ion{Cd}{I} line. While this line is blended with the $\lambda2288$ \ion{As}{I} line, the signature of As is negligible in the spectrum of \thisStarShort; the Cd-As absorption feature could be fit best by just the Cd abundance. Figure \ref{fig:synFits} shows the spectral synthesis fit to the Cd line. 
\par
For In, we determined a 3$\sigma$ upper limit using the \ion{In}{II} line at $2306.06$ \AA. Unfortunately, it is blended with a \ion{Fe}{II} line at $2306.17$ \AA\ with an uncertain \loggf-value \citep{Roederer22} and features a very weak absorption signature in the spectrum of \thisStarShort. For Sn, we detected a possible signature at the $\lambda2287$ \ion{Sn}{I} line, however, it was too noisy to yield a reliable abundance or upper limit. 

\subsection{Second $R$-Process Peak: Te}
The \ion{Te}{I} line at $2385.79$ \AA\ is not resolved in the spectrum of \thisStarShort\ and features a very weak absorption signature (see Figure \ref{fig:synFits}). Furthermore, the neighboring lines at $2385.59$ \AA\ and $2385.92$ \AA, which are possibly \ion{Fe}{I} lines or at least have contribution from them, are not fit well and are blended with the \ion{Te}{I} line. There is additional contribution to the \ion{Te}{I} feature from a \ion{Cr}{I} line at $2385.74$ \AA. Given the difficulty to reliably constrain the contribution from Te in this spectral region, we determined a 3$\sigma$ upper limit of \eps(Te)$\ < 1.83$ for the Te abundance. We show the upper limit model for Te in Figure \ref{fig:synFits}.

\subsection{Ba, Lanthanides, and Hf}
For elements in this group, we determined abundances for all, except for Sm, Tb, and Hf, for which we could only determine upper limits. 
\par
For Ba, we used 5 \ion{Ba}{II} transition lines in the optical, which were strong and provided a precise mean abundance. For La, the line strengths were relatively weaker, but 5 \ion{La}{II} transitions were strong enough to obtain reliable abundance determinations. For Ce, we used two \ion{Ce}{II} lines to determine the abundance. In general, the other lines were either too weak and/or unresolved, and the contribution from Ce was difficult to constrain. For Pr, we determined the abundance using the \ion{Pr}{II} line at 4225.32 \AA. 
\par
For Nd, we determined the abundance using 6 \ion{Nd}{II} lines in the optical. For Sm, unfortunately, the strongest lines were still weak based on our detection threshold (see Section \ref{sec:detLimit}). Therefore, we obtained only a $3\sigma$ upper limit on the Sm abundance using the \ion{Sm}{II} line at $4329.02$ \AA. For Eu, we used 5 \ion{Eu}{II} lines in the optical, all of which lend Eu abundance within $\pm0.10$ dex of each other. 
\par
For Gd abundance, we used two \ion{Gd}{II} lines in the optical and one \ion{Gd}{II} line in the UV. All three lines lend Gd abundance within $\pm0.05$ dex of each other. The $\lambda3032$ \ion{Gd}{II} line in the UV is blended with a \ion{Sn}{I} line at $3032.79$ \AA\ and a \ion{Cr}{I} line at $3029.16$ \AA. The \ion{Cr}{I} line is the dominant source of absorption in this Sn-Gd-Cr feature, but mainly affects the red wing, while the Gd line affects the blue wing. The Sn line is also situated blue and has a direct impact on the Gd abundance. Since the \ion{Sn}{I} lines are extremely weak, if not undetectable in the spectrum of \thisStarShort, we neglect the Sn contribution to this feature (see Section \ref{sec:betweenpeaks}). Furthermore, we added an additional $\pm0.15$ dex component to the systematic uncertainty of this line to compensate for the uncertain contribution of Sn. 
\par
For Tb, we detected a weak absorption signature at the $3874.17$ \AA\ \ion{Tb}{II} line, but its signature is below our detection threshold (Section \ref{sec:detLimit}). As a result, we determined a $3\sigma$ upper limit for the abundance. For Dy, Ho, Er, Tm, and Yb we used 5 \ion{Dy}{II}, 2 \ion{Ho}{II}, 2 \ion{Er}{II}, one \ion{Tm}{II}, and one \ion{Yb}{II} absorption lines, respectively, in the optical spectrum. 
\par
For Lu, we used the \ion{Lu}{II} line at $2615.14$ \AA\ in the UV spectrum, which displayed a strong signature. The HFS pattern for the line was adopted from \cite{DenHartog2020_Lu}. The \ion{Lu}{II} lines in the optical did not display any detectable absorption. While the $\lambda2615$ UV \ion{Lu}{II} is strong, \cite{Roederer2022b} identified that this line is blended with an unknown transition. They showed that this blend primarily affected stars with low levels of \rproc\ enhancement, and would result in an unusually high \eps(Lu/Eu) ratio of $\gtrsim0.5$. Since \thisStarShort\ is highly \rproc\ enhanced, with [Eu/Fe]$\ = +1.37$, we suspect that the unidentified blend, if present, will have a marginal impact on the Lu abundance. Moreover, upon fitting the Lu feature without accounting for the blend, we obtained \eps(Lu)$\ = -0.70$ and \eps(Lu/Eu)$\ = -0.35$. \cite{Roederer22} determined a similar ratio of \eps(Lu/Eu)$\ = -0.40$ for HD 222925 using other \ion{Lu}{II} lines. Nevertheless, for caution, we added an additional $\pm0.10$ dex of systematic uncertainty on the derived Lu abundance. 
\par
We checked for several \ion{Hf}{II} lines in the UV and optical spectra, but could not determine a reliable absorption signature. Therefore, we determined an upper limit of \eps(Hf)$\ < +0.67$. 

\subsection{Third $R$-Process Peak: Os, Pt, and Au}
We determined the Os abundance using the \ion{Os}{II} line at $2282.28$ \AA\ in the UV spectrum. The line is situated between an \ion{Fe}{I} line at $2281.99$ \AA\ and a \ion{Ti}{I} line at $2282.43$ \AA. We adjusted the oscillator strength of these neighboring lines to fit their lines, but we note that these changes did not impact the fit to the Os line and its derived abundance. We used the \loggf\ value for the Os line from \cite{Quinet2006_Os}. However,  \cite{Ivarsson2004_Os} also provided the \loggf\ value for this line, reporting a value lower by $\sim -0.10$ dex. Therefore, we added a $\pm0.10$ dex systematic uncertainty on the Os abundance from the uncertainty in its \loggf\ value.
\par
For Ir, we determined a 3$\sigma$ upper limit using the \ion{Ir}{I} at $2639.71$ \AA\ in the UV, since the signature of the line was weak and unresolved. 
\par
We determined the Pt abundance using two \ion{Pt}{I} lines in the UV at $2659.43$ \AA\ and $2997.96$ \AA. We suspect that there may be an unidentified blend to the $\lambda2997.96$ line (see also Figure 4 of \citealt{DenHartog2005_PtI}), especially since the abundance from this line is higher by $0.15$ dex than the abundance from the $\lambda2659.43$ line. Therefore, we included an additional $\pm\ 0.10$ dex systematic uncertainty on this line abundance. The atomic parameters for both the lines, including HFS and IS, were taken from \cite{DenHartog2005_PtI}. While we detected absorption signatures at some other \ion{Pt}{I} lines listed in \cite{DenHartog2005_PtI}, the lines were either not well-resolved or weak. 
\par
For the Au abundance, we used the \ion{Au}{I} resonance lines at $2427.95$ \AA\ and $2675.95$ \AA. The $\lambda2427.95$ line is blended with an \ion{Fe}{II} line at $2428.08$ \AA, which contributes to the red side of the Au-Fe absorption signature, but constrains the data well. The \loggf\ value for this \ion{Au}{I} line was taken from \cite{Hannaford1981_gold}, with NIST ASD quoting a grade B+ (7\%, $\pm\ 0.05$ dex uncertainty). For the $\lambda2676$ \ion{Au}{I} line, we used the HFS line component provided by \cite{Roederer22}. The \loggf\ value of the line was also taken from \cite{Hannaford1981_gold}, but for this line, NIST ASD quotes a higher accuracy of A+ (2\%, $\pm\ 0.02$ dex uncertainty). In our spectrum, this line is primarily blended with an unidentified line. Since the resolution of our spectrum is not high enough to resolve the blend from the \ion{Au}{I} line, we used the atomic parameters of this unidentified blend as constrained by \cite{Roederer22} for HD 222925. This \ion{Au}{I} line is also blended with a \ion{Nb}{II} line at $2675.94$ \AA, for which we estimated a negligible contribution, since the Nb lines in \thisStarShort\ appear almost undetectable. As a final note, we blue-shifted the wavelength of the $\lambda2676$ Au line component by $0.04$ \AA\,\ from the original center-of-gravity wavelength reported in NIST. The need for a similar wavelength shift was discussed in \cite{Roederer22} for HD 222925, but the cause is unclear. 

\subsection{Actinides}
We did not detect any signature of Th or U at any of the \ion{Th}{II} and \ion{U}{II} lines, including the recently analyzed \ion{U}{II} lines at $4050.04$ \AA\ and $4090.23$ \AA\ \citep{Shah2023}. Therefore, we obtained 3$\sigma$ upper limits for the Th abundance. We do not report the upper limit for U since the spectral regions of the lines were noisy, preventing a constraining upper limit determination.

\begin{figure*}
    \centering
    \includegraphics[width = 0.9\textwidth]{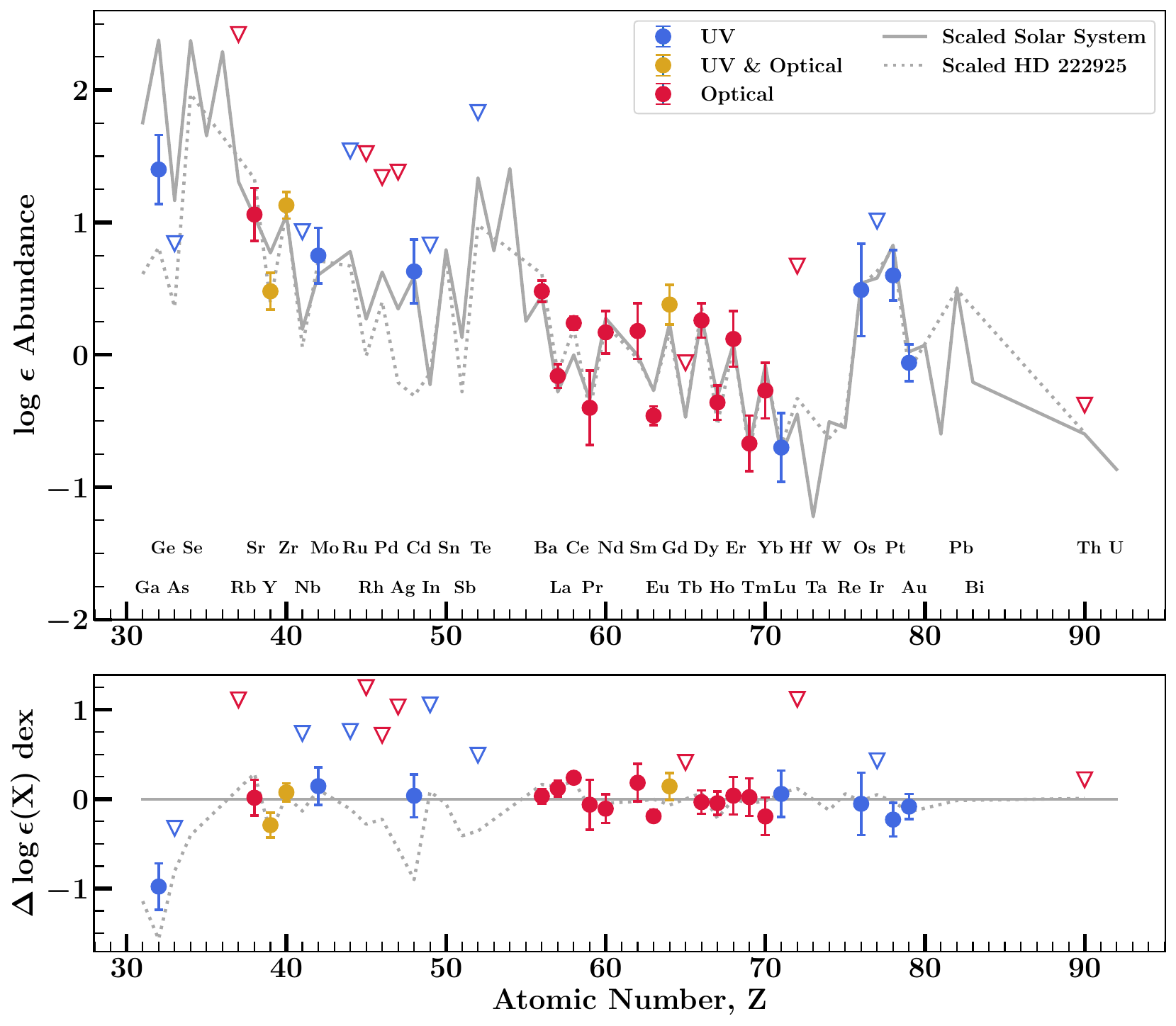}
    \caption{Adopted abundances of the \rproc\ elements for \thisStarShort\ shown with different colored data points, depending on the wavelength-region used for deriving the abundance. Abundances obtained with the UV spectrum are shown in blue, with optical spectrum are shown in red, and with both UV and optical are shown in yellow. Upper limits are shown in downward-triangle with the same color map as the abundances. The grey-solid trace depicts the scaled S.S. \rproc\ pattern as obtained from \protect \cite{Prantzos2020_SS}. The grey-dotted line depicts the scaled \rproc\ pattern of the RPE star, HD 222925, as obtained from \protect \cite{Roederer22}}
    \label{fig:rprocessPattern}
\end{figure*}

\section{Detection Threshold and Upper Limits}\label{sec:detLimit}
Due to the high temperature of the star, several absorption lines of the \rproc\ elements are weak and exhibit absorption depths comparable to the noise level of the spectrum. In order to enable a reliable abundance determination from these weak lines, we employed a minimum EW (EW$_{limit}$) as the detection threshold. We obtained EW$_{limit}$ using the Cayrel formula for uncertainty on the EW of a weak line ($\sigma_{EW}$), as given by equation \ref{cayrel_formula} \citep{Cayrel1988, Cayrel2004}. Here $\delta$x is the width of the pixel in \AA, which is $\sim0.04$ \AA\ for both the UV and optical spectrum. We then obtained EW$_{limit}$ as 2$\sigma_{EW}$, as shown in equation \ref{eqn:EWLimit}.
\begin{equation}\label{cayrel_formula}
    \mathrm{\sigma}_{EW} = \frac{1.5}{S/N}\sqrt{FWHM*\delta x}
\end{equation}
\begin{equation}\label{eqn:EWLimit}
    \mathrm{EW}_{limit} = 2\sigma_{EW}
\end{equation}
With this method, the EW$_{limit}$ is 1.6 m\AA\ for the UV spectrum and 3.8 m\AA\ for the optical spectrum. Therefore, we only use 
lines that have EWs, as obtained from the best-fit spectral synthesis model, larger than the EW$_{limit}$ of the respective spectrum. In general, most of the weak lines used for abundance determination have EWs larger than $3\sigma_{EW}$. One notable exception is the $\lambda3039$ \ion{Ge}{I} line, which has an EW measurement of 2.1$\sigma_{EW}$. We further discuss the abundance determination with this line in Section \ref{sec:disc_lightRprocess}. In addition to this detection threshold, we take into account unknown blends and the overall fit to the spectral region when using a line for abundance determination. 
\par
For elements that did not have any lines with signatures beyond the detection threshold, we determined a 3$\sigma$ upper limit on the element abundance with a suitable line of the element. The most suitable line for upper-limit determination was chosen based on the strength of the line, the SNR in the spectral region, and the constraint on the neighboring lines. Specifically, the $3\sigma$ upper limit was obtained such that the least $\chi^{2}$ value of the upper-limit spectral-synthesis model was higher than the least $\chi^{2}$ value of the best-fit spectral-synthesis model by $3\sigma$, where $\sigma$ is the spectral noise of the data. 

\section{Uncertainty Analysis}\label{sec:uncertainty}
To determine the total uncertainty ($\sigma_\mathrm{tot}$) on the abundance of an element, we took into account statistical uncertainties ($\sigma_\mathrm{stat}$), which represented uncertainties in the determination of the atomic parameters of the lines, and systematic uncertainties ($\sigma_\mathrm{sys}$), which represented uncertainties in the stellar parameters and line fits. These are listed in Table \ref{tab:adopted_abund}. We added the $\sigma_\mathrm{stat}$ and $\sigma_\mathrm{sys}$ uncertainties in quadrature to obtain $\sigma_\mathrm{tot}$.
\par
For $\sigma_\mathrm{stat}$, we used the standard deviation in the abundances from individual lines, in the cases 6 or more lines were used ($N \geq 6$). In case, $2 \leq N \leq 5$, we obtained  $\sigma_{stat}$ by multiplying the range of the abundances with the $k$-factor from \cite{keeping1962}, to compensate for the small samples and obtain more realistic uncertainty estimates \citep[e.g.,][]{Cain2018}. In the case $N = 1$, we assigned a fiducial uncertainty of $\pm0.20$ dex to $\sigma_\mathrm{stat}$. 
\par
For $\sigma_{sys}$, we accounted for uncertainties from stellar parameters, \eTeff, \logg, \vmicro,  and [Fe/H] and additional uncertainties from spectral-synthesis fitting due to blends, uncertain atomic parameters, or uncertain continuum placement (see Section \ref{sec:chemAbund} for details). We first obtained $\sigma_{sys}$ for each line and then took their average, $<\sigma_{sys}>$,  to represent the systematic uncertainty on the mean abundance of the element, which we list in Table \ref{tab:adopted_abund}. To obtain the uncertainties from the stellar parameters, we individually changed \eTeff, \logg, \vmicro, and [Fe/H] by $\pm 82$ K, $\pm\ 0.07$ dex, $\pm\ 0.2$ km/s, and $\pm\ 0.2$ dex, respectively. We added added the uncertainties from each of the stellar parameters as well as the uncertainty from the spectral models in quadrature to obtain $\sigma_{sys}$ for each line. 
\par
For the final systematic uncertainty of a species, we used the average of the systematic of species' transitions. The statistical and systematic uncertainties for the adopted abundances are listed in Table \ref{tab:adopted_abund}. 

\section{Discussion}\label{sec:Discussion}
\subsection{Light Elements}  
Among the light elements with Z $\leq30$, we obtained abundances for 16 elements, including 26 species. In the following sections, we discuss the abundances from the UV and optical lines (section \ref{sec:uvop_disucss_light}), the observed NLTE effects and theoretical corrections (section \ref{sec:NLTE_discuss_light}), and  the adopted abundances of \thisStarShort\ compared to that of other metal-poor stars from \cite{Roederer2014_largesample} (section \ref{sec:litcompare_discuss_light}). 

\subsubsection{UV and Optical Abundances}\label{sec:uvop_disucss_light}
The UV spectral coverage enabled us to determine abundances of 4 unique species, which are generally not detectable in the optical spectra, including \ion{Mg}{II}, \ion{Al}{II}, \ion{Co}{II}, and \ion{Ni}{II}. Additionally, since these are the dominant species,  their detection enabled an empirical test of the NLTE effects affecting the corresponding neutral species of these elements (see Section \ref{sec:NLTE_discuss_light}). 
\par
The UV spectral coverage also benefited the abundance determinations of 14 species, increasing the number of absorption lines available.
These species included \ion{Mg}{I}, \ion{Si}{I}, \ion{Si}{II},  \ion{Sc}{II}, \ion{Ti}{II}, \ion{V}{II}, \ion{Cr}{I}, \ion{Cr}{II}, \ion{Mn}{I}, \ion{Mn}{II}, \ion{Fe}{I}, \ion{Fe}{II}, \ion{Co}{I}, and \ion{Ni}{I}. We find that the optical and UV abundances of these species agree well, within uncertainties. This can be seen in the top panel of Figure \ref{fig:Fe_groupElements}, which shows the individual optical (black data) and UV (green data) [X/Fe] abundance ratio of each species. The agreement in the UV and optical abundances of these species validates our reduction and analysis techniques for the two spectra \citep[e.g.,][]{Roederer2022b}. 
\par
On the other hand, optical spectral coverage enabled us to determine abundances for 8 species which were not available in the UV, including \ion{O}{I}, \ion{Na}{I}, \ion{Al}{I}, \ion{K}{I}, \ion{Ca}{I}, \ion{Ti}{I}, \ion{V}{I}, and \ion{Zn}{I}. 

\subsubsection{Neutrals, Ions, and NLTE Effects}\label{sec:NLTE_discuss_light}
We obtained abundances of both the neutral and first-ionized species for several light elements. We plot these abundances as [X/Fe] abundance ratio in the middle panel of Figure \ref{fig:Fe_groupElements}, with square data points for neutral species and circular data points for first-ionized species. For this plot, we combined the UV and optical abundances of the species; both the UV and optical lines are considered for the mean [X/Fe] abundance ratio and the corresponding uncertainty of the species following the method described in Section \ref{sec:uncertainty}. We also depict the LTE and NLTE-corrected abundances for each species separately, colored-coded with black and red data points, respectively. 
\par
As seen in Figure \ref{fig:Fe_groupElements}, the abundances of the neutral and ionized species of most light elements agree. On the other hand, for HD 222925, \cite{Roederer22} observed that the abundances of the neutral species of Ca and Fe-group elements were all systematically lower than the abundances of their ionized counterparts. They noted that these offsets were consistent with over-ionization effects predicted for these species due to LTE assumptions. Given the higher \logg\ of \thisStarShort, we suspect that the over-ionization effects are lower in this case and/or the precision of the abundances derived here is not sufficient to discern any systematic offset between the abundances of the neutral and ionized species.
\par
We also find that the NLTE-corrected abundances of the neutral species (red-square points in the middle panel of Figure \ref{fig:Fe_groupElements}) generally agree well with the LTE abundances of the corresponding ionized species (black circular points).  The agreement is very good for Mg and Si.  In the cases of \ion{Mg}{I} and \ion{Si}{I}, the NLTE corrections are also small i.e., $+0.10$ dex and $+0.00$ dex, respectively. For Ti, Cr, Mn, and Fe, the agreement exists within uncertainties. We note that the NLTE corrections for \ion{Cr}{I} ($+0.25$ dex), \ion{Mn}{I} ($+0.28$ dex), and \ion{Fe}{I} ($+0.17$ dex) are at least 1$\sigma$ higher than their respective uncertainties. Such significant corrections highlight the importance of NLTE studies and accounting for NLTE corrections as we move towards larger and and more precise spectroscopic surveys and analyses. Overall, the general agreement between the NLTE-corrected abundances of the neutral species and the LTE abundances of the corresponding ionized species, in spite of the large corrections, reflects well on the current state of NLTE theoretical models for most species. Although, as spectroscopic studies start achieving higher precision in the abundances, discrepancies might be revealed.
\par
An important exception to this case is Al, wherein the NLTE-corrected abundance of \ion{Al}{I} and the LTE abundance of \ion{Al}{II} are discrepant by $\sim0.4$ dex. The NLTE correction for the $\lambda3961$ \ion{Al}{I} resonance line is significant and is estimated to be $+0.47$ dex by \cite{Mashonkina2016_AlSiNLTE}. We also used the NLTE grids from \cite{Nordlander2017_AlNLTE}\footnote{\url{https://www.mso.anu.edu.au/~thomasn/NLTE/data/}}, to obtain a similar correction of $+0.49$ dex. Other NLTE studies, such as \cite{Lind2022_NaMgAlNLE}, suggest even smaller NLTE corrections for \ion{Al}{I}, on the order of $\sim0.20$ dex. Interestingly, even with this significant correction, the NLTE-corrected abundance of \ion{Al}{I} using the $\lambda3944$ and $\lambda3961$ resonance lines is still $\sim0.40$ dex smaller than the \ion{Al}{II} LTE abundance with the $\lambda2669$ line. Moreover, \cite{Mashonkina2016_AlSiNLTE} and \cite{Nordlander2017_AlNLTE} showed that NLTE effects for \ion{Al}{II} are negligible. Therefore, for this study we adopted the \ion{Al}{II} abundance. However, we recommend future NLTE studies to to investigate the source of this discrepancy.
\par
As noted in Section \ref{sec:chemAbund}, for Li, O, Na, K, Ca, and Zn, we adopted the NLTE-corrected abundance of the neutral species (red square points in the middle panel of Figure \ref{fig:Fe_groupElements}), since the abundances of their ionized species were not available. The NLTE corrections for some of these elements are significant e.g., $-0.28$ dex for \ion{K}{I}, $+0.14$ dex for \ion{Ca}{I}, and $+0.18$ dex for \ion{Zn}{I} (also see Table \ref{tab:UVOP_abund}), indicating that it is especially important to take NLTE corrections into account for these elements, since their ionized species are generally not available. On the other hand, we note that the analysis of \ion{Zn}{I} and \ion{Zn}{II} abundances by \cite{RoedererBarklem2018} for metal-poor dwarf and subgiant stars indicated minimal ($\lesssim\ 0.10$ dex) departures from LTE.
\par
\subsubsection{Comparison to Other Metal-Poor Stars}\label{sec:litcompare_discuss_light}
We show the adopted light-element abundances of \thisStarShort\ in the bottom panel of Figure \ref{fig:Fe_groupElements} with black data points and in the form of [X/Fe] abundance ratios. For comparison, we also display the typical values and range of the [X/Fe] abundance ratios for metal-poor stars. For this purpose, we used the light-element abundance ratios of $247$ stars, as part of a sample of $313$ stars analyzed by \cite{Roederer2014_largesample}. We obtained the abundances using \texttt{JINAbase}\footnote{\url{https://jinabase.pythonanywhere.com/index}} \citep{Abohalima2018_JINA}. Specifically, we show the mean [X/Fe] abundance ratios of this sample with a grey-solid line and the $\pm1$ standard deviation from the mean with a shaded-grey region. As seen in the figure, we have obtained abundances of several light elements for \thisStarShort.
\par
The resulting [X/Fe] abundance ratios of \thisStarShort\ generally compare well with that of the metal-poor stars in the \cite{Roederer2014_largesample} sample. Some discrepancies include [Al/Fe] and potentially [Ca/Fe], due to inconsistency in the adopted values.  For [Al/Fe], we adopted the \ion{Al}{II} abundance, which is much higher than the \ion{Al}{I} abundance due to NLTE effects (see Section \ref{sec:NLTE_discuss_light}), whereas \cite{Roederer2014_largesample} adopted the \ion{Al}{I} abundance. Similarly, for [Ca/Fe], we adopted the NLTE-corrected \ion{Ca}{I} abundance, while \cite{Roederer2014_largesample} adopted the LTE \ion{Ca}{I} abundance, which is generally slightly lower. We note that for \ion{O}{I}, \ion{Na}{I}, and \ion{K}{I}, the NLTE-corrected abundances were adopted by both this study and \cite{Roederer2014_largesample}. Our final note is that for Ti, V, Cr, and Mn, both this study and \cite{Roederer2014_largesample} adopted the abundance of the ionized species; on the other hand, for Mg, Si, and Co, we adopted the abundance of the ionized species and \cite{Roederer2014_largesample} adopted the abundance of the neutral species, but no difference is observed in the abundance ratios. 
\par
In general, the trends in the light-element abundances of \thisStarShort\ compare well with that of other metal-poor stars, indicating that the primary enrichment channel for the light elements of \thisStarShort\ was core-collapse supernovae. We obtained an $\alpha$-enhancement of [$\alpha$/Fe] $= +0.42$ for \thisStarShort, using abundances of O, Na, Mg, Si, and Ca, which is typical of metal-poor stars \citep{Roederer2014_largesample, Cowan2020_FePeakElements}. We included Ti as an iron-group element here, given the discussions in \cite{curtis2019} and \cite{Cowan2020_FePeakElements}. The iron-group elements also exhibit the typical trends, including enhancement of Sc, Ti, and V, followed by solar ratios for Cr, Co, and Ni, followed by some enhancement for Zn \citep{Cowan2020_FePeakElements, Sneden2023_FePeak}. 
\par
The elemental abundances not shown in Figure \ref{fig:Fe_groupElements} are Li, CH, and CN. The Li abundance of \eps(Li)$= 2.35$, including the NLTE correction, falls along the Spite Plateau \citep{Spite1982, Norris2023}, which is expected for a turn-off star. Finally, with [C/Fe] $= +0.62$, using the CH abundance, we confirm that the star is not carbon enhanced. All in all, we determined abundances for 16 elements and one molecule (CH), and a 3$\sigma$ upper limit for one element (S) and one molecule (CN).

\subsection{Heavy Elements and the $R$-Process Pattern}
\begin{figure}
    \centering
    \includegraphics[width = 0.49\textwidth]{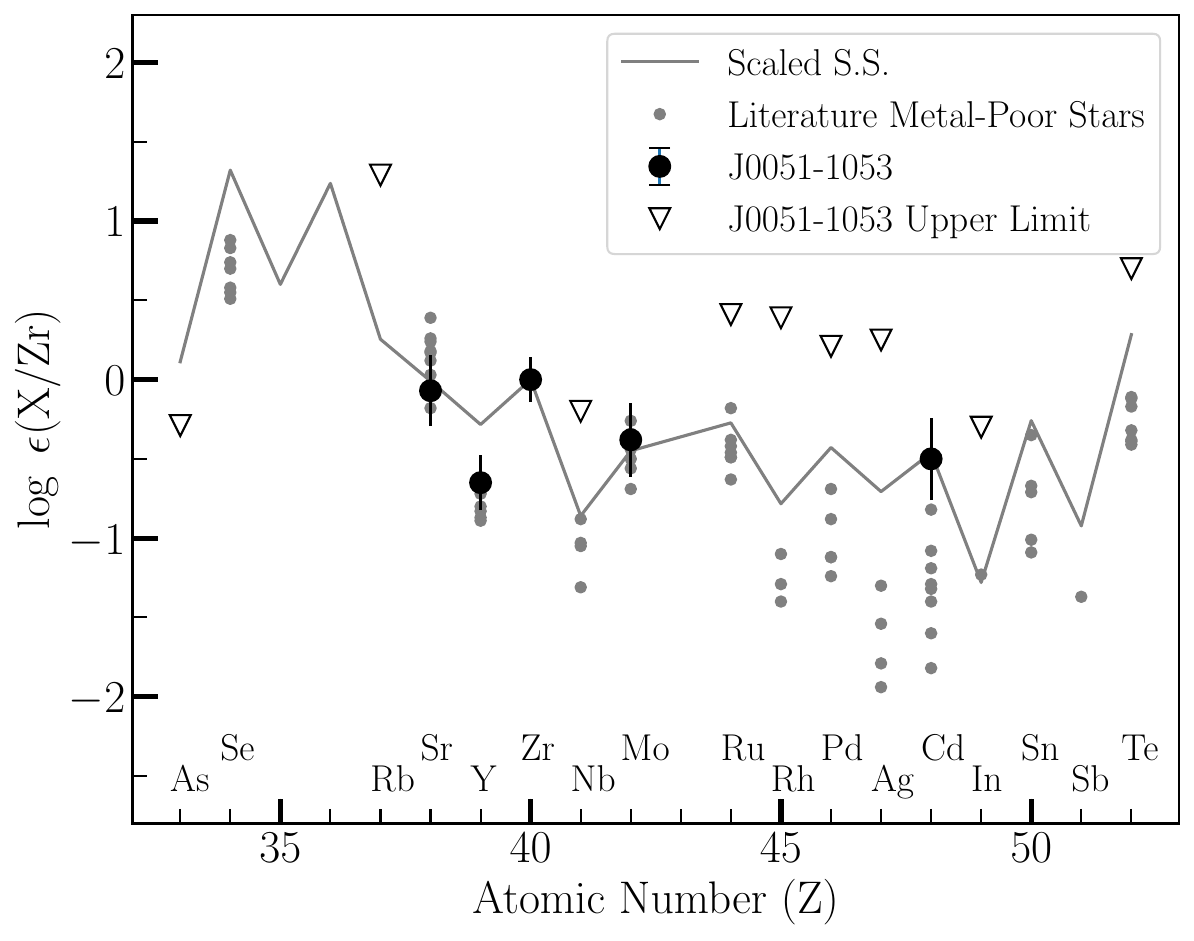}
    \caption{Light \rproc\ elements of metal-poor stars scaled to Zr. As identified by \protect \cite{Roederer2022b}, abundances for Se, Sr, Y, Zr, Nb, Mo, and Te show minimal dispersion for the 8 metal-poor stars in their sample (grey data points). The S.S. \rproc\ scaled to Zr is shown with a grey-solid line. We note that the scaled Sr, Y, and Mo abundances of \thisStarShort\ (black data points) follow the trend of other metal-poor stars. On the other hand, the scaled Cd abundance of \thisStarShort\ is much higher, the highest observed in metal-poor stars so far, adding to the observed dispersion of Cd abundances in metal-poor stars.}
    \label{fig:zrscaled}
\end{figure}

\begin{figure*}
    \centering
    \includegraphics[width = 0.9\textwidth]{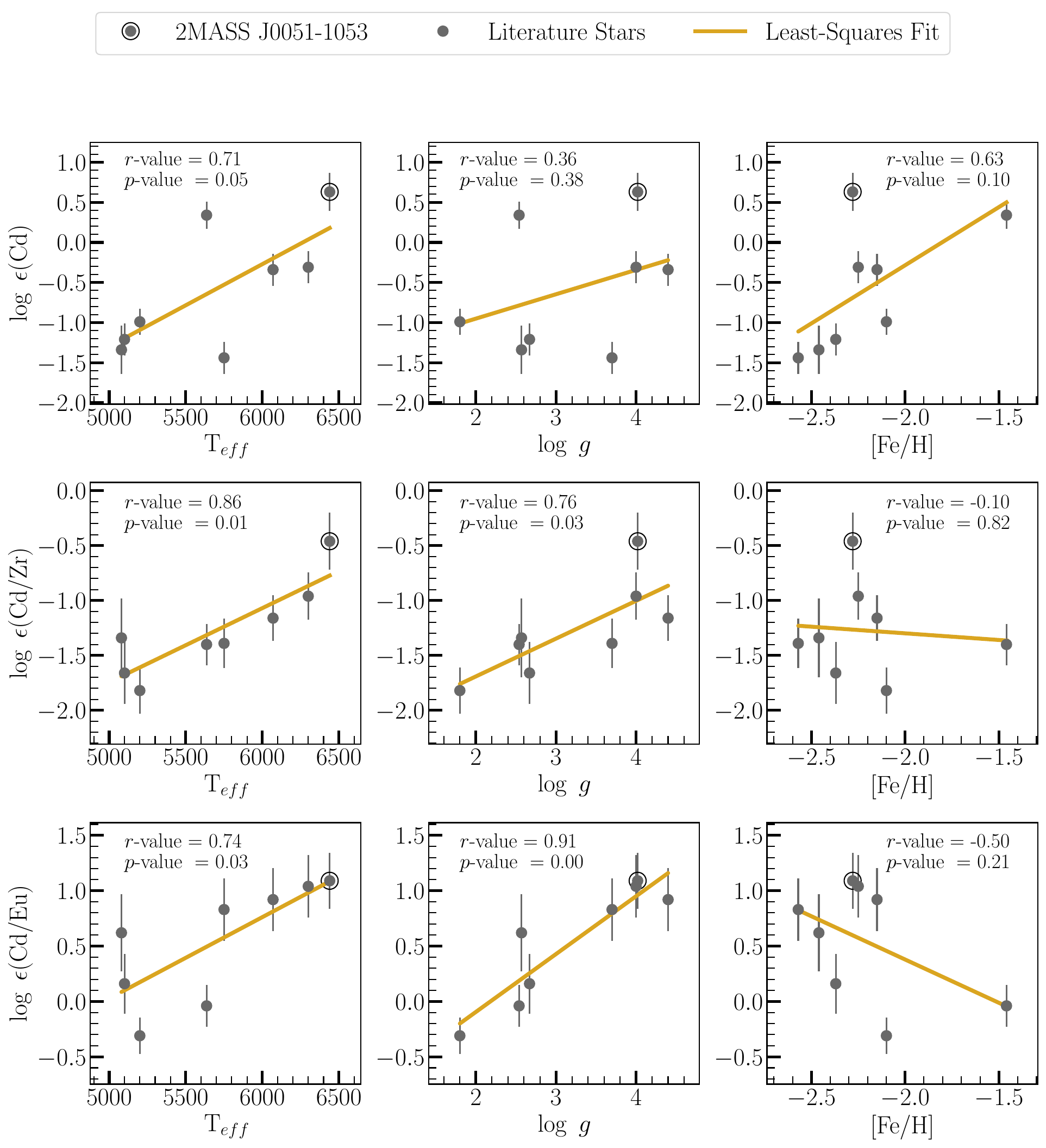}
    \caption{Cd abundances of \thisStarShort\ and other metal-poor stars \protect \citep[][and references therein]{Roederer2022b} with respect to stellar parameters of the stars, including \eTeff, \logg, and [Fe/H]. The absolute Cd abundances are shown in the top row, Cd abundance scaled to Zr are shown in the middle row, and Cd abundance scaled to Eu are shown in the bottom row. Also shown is the least-square fit to the data with a golden-solid line. The Pearson correlation coefficient (\textit{r}-value) and the corresponding probability of observing a correlation of at least this magnitude by chance (\textit{p}-value) are listed on the top for each panel.}
    \label{fig:Cd}
\end{figure*}

We obtained abundances for 23 \rproc\ elements with  31 $\leq$ Z  $\leq$ 92.  Out of these 23 elements, we obtained abundances of 8 elements with the UV spectrum and 3 elements with both the UV and optical spectrum. We also obtained $3\sigma$ upper limits on the abundances of 13 \rproc\ elements. The UV spectrum especially enabled the upper limit determination for 6 of these elements. 
\par
We show the final abundances of all the neutron-capture elements in Figure \ref{fig:rprocessPattern}. Given the \rproc\ enhancement of the star as determined by [Eu/Fe]$ = +1.37$ and [Ba/Eu] $= -0.72$ \citep{Beers&Christlieb2005Rev, Frebel18_rev}, we confirm the findings of other studies \citep{RPA3, Gull2021_helmiStream} that \thisStarShort\ is an $r$-II star i.e., a highly \rproc-enhanced star with [Eu/Fe]$ > +0.3$. This classification is further affirmed by the abundance pattern of the neutron-capture elements, which generally matches the S.S. \rproc\ pattern, as seen in Figure \ref{fig:rprocessPattern}. Here the S.S. \rproc\ pattern is scaled to the \rproc\ pattern of \thisStarShort\ using the mean lanthanide abundance. For the mean lanthanide abundance, we specifically used abundances of elements from Ba to Hf, except for Sm and Tb, since we only have upper limits on their abundances. For comparison, we also show the abundance pattern of HD 222925, a metal-poor highly \rproc-enhanced star with abundances of 42 \rproc\ elements determined by \cite{Roederer2018_HD222925} and \cite{Roederer22}. We note that the data points for \thisStarShort\ in the Figure \ref{fig:rprocessPattern} are color-coded based on the origin of the abundance of the elements from the UV spectrum (blue), optical (red), or both (yellow). We further discuss the \rproc\ pattern of \thisStarShort\ and compare it to the S.S. and HD 222925 \rproc\ patterns below.

\subsubsection{$R$-Process Pattern from the First to the Second $R$-Process Peak}\label{sec:disc_lightRprocess}
Various studies have indicated that the abundance pattern of the elements in this region (from Z = 31 to Z = 52) have a scatter and deviate from the S.S. \rproc\ pattern, even in RPE stars \citep[e.g.,][]{SiqueiraMello2014_rI, Ji2016_RetII_APJ}. The origin of this effect is still unknown, with different astrophysical sites, conditions, and processes being considered \citep[e.g.,][]{Chiappini2011, Hansen2012, Wanajo2013, Holmbeck2019}. For \thisStarShort, we also observe the Ge and Y abundances to be lower than the scaled S.S. \rproc\ pattern (see Figure \ref{fig:rprocessPattern}). Additionally, the upper limit on the As abundance is also significantly lower than the scaled S.S. \rproc\ pattern. On the other hand, we observe the Sr, Zr, Mo, and Cd abundances to follow the scaled S.S. \rproc\ pattern. 
\par
In the case of Ge, we find that its abundance ratio of [Ge/Fe]$\ = +0.10$ is significantly higher than observed in $\sim20$ metal-poor stars so far, which have a mean [Ge/Fe] of $\sim\ -0.90$ with a standard deviation of $\sim0.26$, indicating a difference in the origin \citep[][and references therein]{Cowan2005_largeSample, Roederer2014_AsSe, Peterson+2020}. {\cite{Cowan2005_largeSample} showed that for metal-poor stars, [Ge/H] ratio is correlated with [Fe/H], pointing to a common origin of Ge and Fe-peak elements, possibly $\alpha$-rich freezeout in core-collapse supernovae (also see \citealt{Roederer2014_AsSe} for a similar discussion and larger sample of abundances). However, other mechanisms are also capable of producing Ge, including the $\nu p$-process \citep[e.g.,][]{Frohlich2006} and the \rproc\ \citep[e.g.,][]{Farouqu2010_highentropywind}, both of which have multiple potential astrophysical sites. Therefore, the high [Ge/Fe] ratio observed here may indicate a production mechanism of Ge decoupled from that of Fe. As noted by \cite{Roederer2014_AsSe}, it will be challenging to establish the originating mechanism without the isotopic abundances. Nevertheless, this result renews the need for more Ge abundance determinations in metal-poor stars. We also consider that the Ge abundance derived here may not be dependable, since only one \ion{Ge}{I} was used. Additionally, there could be NLTE effects present. Therefore, caution is warranted before over-interpreting this Ge abundance. 
\par
Interestingly, in a recent study, \cite{Roederer2022b} showed using 8 metal-poor stars, including 6 RPE, that the relative abundances of Se, Sr, Y, Zr, Nb, Mo, and Te in these stars actually all agree, provided that the abundances are scaled by a light \rproc\ element. We find \thisStarShort\ to conform to this rule based on the derived abundances of Sr, Y, Zr, and Mo. We show this agreement in Figure \ref{fig:zrscaled}, where we have plotted the Zr-scaled abundances of the \cite{Roederer2022b} sample stars with grey data points, and the Zr-scaled abundances of \thisStarShort\ with black data points. We also show the Zr-scaled \rproc\ pattern of the S.S. from \cite{Prantzos2020_SS} with a grey-solid line. Additionally, although we only have $3\sigma$ upper limits for Nb and Te for \thisStarShort, they don't rule out a possible agreement with other metal-poor stars. In fact, we note that the expected Te abundance, based on the Zr-scaled S.S. \rproc\ pattern lends a stronger absorption signature for the \ion{Te}{I} line than the observed data suggests, indicating that the Zr-scaled Te abundance for \thisStarShort\ is not as high as the Zr-scaled S.S. Te abundance. On the other hand, the expected Te abundance, based on the Zr-scaled Te abundance of HD 222925 is more likely to be a better fit to the observed data. Therefore, our current results for \thisStarShort\ supports the hypothesis proposed by \cite{Roederer2022b} that the enrichment source of the light \rproc\ elements in the early Universe must have been common and consistent in producing the observed relative abundances of these elements. 
\par
We note here that several studies have pointed out the significant deviation of Y in metal-poor stars from the scaled S.S. \rproc\ pattern \citep[e.g.,][]{Sneden2003_CS22892, Roederer2022b}, as can be seen in Figures \ref{fig:rprocessPattern} and \ref{fig:zrscaled}. A recent study by \cite{Storm2023} computed NLTE departure coefficients for the low- and high-excitation \ion{Y}{II} lines for a range of stellar parameters and found that these corrections can be as large as $+0.50$ dex for low-excitation \ion{Y}{II} lines in metal-poor red-giant stars. For a metal-poor turn-off star, they calculated NLTE corrections on the order of $\sim+0.15$ dex for the low-excitation \ion{Y}{II} lines. They proposed that such NLTE effects could contribute to the observed deviation of Y abundances from the S.S. \rproc\ pattern. Indeed, we find that taking into account a $+0.15$ dex of NLTE effects for the low-excitation \ion{Y}{II} lines used here would enable an agreement between the lanthanide-scaled Y abundance of \thisStarShort\ and the S.S. \rproc\ pattern within uncertainties.
\par
Although, as seen in Figure \ref{fig:zrscaled}, such an agreement in the relative abundances of the light \rproc\ elements in metal-poor stars does not extend to Ru, Rh, Pd, Ag, Cd, and Sn. Of these elements, for \thisStarShort, we have abundance determined for Cd and $3\sigma$ upper limits determined for Ru, Rh, Pd, and Ag. We find that \thisStarShort\ has the highest reported \eps(Cd), \eps(Cd/Zr) and \eps(Cd/Eu) abundance among the 9 metal-poor stars with Cd abundance determined so far \citep{Roederer2014_AsSe, Peterson+2020, Roederer22, Roederer2010}. This record-high abundance further contributes to the large dispersion in the \eps(Cd/Zr) abundance ratio of metal-poor stars, as seen in Figure \ref{fig:zrscaled}. Note that we have not included the Cd abundance of HD 196944 from \cite{Placco2015_HST} in Figure \ref{fig:zrscaled}, since its chemical composition mainly originates from the \sproc. 
\par
While the production of Cd has not been of specific focus due to the small number of observations, theoretical studies have indicated that elements in this region (44 < Z < 51) could be formed through the weak \rproc\ i.e., a neutron-capture process with intermediate neutron densities \citep[e.g.,][]{Montez2007}. On the other hand, \cite{Vassh2020} showed that the elements in this region could also be products of late-time fissioning of heavy nuclei, which would also deposit material in the lanthanide region. In that case, the abundances of these elements would be correlated with heavier \rproc\ elements like Eu, instead of other light \rproc\ elements like Zr. Indeed, \cite{Roe2023_science} observed correlations of Ru, Rh, Pd, and Ag abundances of \rproc\ stars with their [Eu/Fe] ratios, supporting this theory and providing an explanation for the observed dispersion in the abundances of these elements when scaled with Zr, as seen in Figure \ref{fig:zrscaled}. However, as shown by \cite{Roe2023_science}, Cd abundances of \rproc\ stars do not show a similar correlation with [Eu/Fe]. We report here that this star adds to the evidence of a large dispersion in the Cd abundance, for which the cause continues to be unknown.
\par
While the cause of such a dispersion of Cd abundances could be astrophysical, here we check for possible causes due to limitations in our stellar atmospheric models. For this purpose, we plot \eps(Cd), \eps(Cd/Zr), and \eps(Cd/Eu) abundances of all metal-poor stars with Cd abundance determined so far against their \eTeff, \logg, and [Fe/H] in Figure \ref{fig:Cd}.  The stars included are HD 128279 \citep{Roederer2014_AsSe}, HD 108317 \citep{Roederer2014_AsSe}, HD 140283 \citep{Peterson+2020}, HD 19445 \citep{Peterson+2020}, HD 84937 \citep{Peterson+2020},  BD +17\degree3248 \citep{Roederer2022b}, and HD 222925 \citep{Roederer22}. We note that we only used \ion{Cd}{I} abundances of these stars since these are typically the reported Cd abundances. We highlight the data points of \thisStarShort\ with a circular boundary around them. 
\par
We also fit the data with a linear least-square regression line, shown with golden-solid lines in Figure \ref{fig:Cd}. We list the Pearson correlation coefficient ($r$-value) and the corresponding $p$-value of the correlation in each panel. The $r$-value indicates the degree to which the two quantities are (anti-)correlated with ($-1$)$+1$ indicating the strongest (anti-)correlation possible and 0 indicating no correlation. The $p$-value indicates the probability of a correlation equal to or stronger than what is observed, given the null hypothesis of no correlation. We find significant correlations ($-0.5 < r$-value $< 0.5$ and $p$-value $< 0.05$) of \eps(Cd), \eps(Cd/Zr), and \eps(Cd/Eu) with respect to \eTeff, and of \eps(Cd/Zr) and \eps(Cd/Eu) abundances with respect to \logg. We find no significant correlation of any Cd-based abundances with respect to [Fe/H]. 
\par
Given the observed trends in the Cd abundance with respect to \eTeff\ and \logg, we consider the case that the formation of the $\lambda2288$ \ion{Cd}{I} line is strongly affected by NLTE. We note that \cite{Peterson+2020} used \ion{Cd}{II} and \ion{Cd}{I} lines for 4 metal-poor stars. While they derived consistent abundances between the two species, they commented that the consistent abundances were only possible by ignoring the HFS of the \ion{Cd}{II}, as reported in \cite{Roederer&Lawlor2012}. In the case that they used the HFS, inconsistent abundances were obtained. Such an inconsistency could further indicate strong NLTE effects for \ion{Cd}{I}. With the present information, it is difficult to speculate the exact dependence of the NLTE effects on \eTeff\ and \logg, if any, since their effects are degenerate. Therefore, we strongly urge the community to carry out detailed calculations to investigate the NLTE effects on \ion{Cd}{I} abundances. We also encourage further theoretical investigations into a possible astrophysical origin for the dispersion in Cd abundances of metal-poor stars. Additional abundance determinations of \ion{Cd}{I} and \ion{Cd}{II} in metal-poor stars will also be helpful. 

\subsubsection{$R-$Process Pattern from the Lanthanides to the Third $R$-Process Peak}
We find that, except for Ce and Eu, the \rproc\ elements of \thisStarShort\ in the lanthanide region, from Ba (Z = 56) to Lu (Z = 71), adhere to the scaled \rproc\ patterns of the S.S. and HD 222925. While the Ce abundance of \thisStarShort\ is higher than the scaled S.S \rproc\ abundance by $0.24$ dex, it agrees well with the scaled Ce abundance of HD 222925. Since we are using only two weak \ion{Ce}{II} lines, we believe that a higher resolution and S/N study is warranted to better understand this discrepancy with the scaled S.S. Ce abundance.
\par 
On the other hand, the Eu abundance of \thisStarShort\ is lower than the the scaled Eu abundances of both the S.S. and HD 222925 by $\sim0.18$ dex, which is more than 1$\sigma$. Since we have used $5$ \ion{Eu}{II} lines, all of which are of intermediate strength and have been well-established in the literature, we find this Eu discrepancy of note. Typically, discrepancies between the Eu abundance of RPE stars and the S.S. \rproc\ pattern are not discovered or reported, since the S.S. \rproc\ patterns are often scaled relative to the Eu abundance of the stars. However, as noted previously, in this study we have scaled the \rproc\ pattern of the S.S. and HD 222925 using the mean lanthanide abundance.
\par
To understand this discrepancy, we consider the effects of NLTE on the formation of the \ion{Eu}{II} lines. \cite{Zhao+2016} estimated that for BD-04 3208, a dwarf metal-poor star with stellar parameters similar to \thisStarShort\ \citep{Sitnova+2015}, the Eu NLTE abundance was $0.13$ dex higher than the LTE abundance, when using the $\lambda4129$ \ion{Eu}{II} line. A similar NLTE correction for the mean Eu abundance of \thisStarShort, which is within $0.01$ dex of the $\lambda4129$ abundance, would enable an agreement of its Eu abundance with the scaled Eu abundances of the S.S. and HD 222295 within uncertainties. 
\par
While there have been several studies indicating that the universal \rproc\ pattern extends to the third \rproc\ peak \citep[e.g.,][]{Cowan2005_largeSample,Roederer2009, Barbuy2011_CS31082, Roederer22}, there have been only $5$ RPE stars with a Au abundance determined \citep{Cowan2002, Sneden2003_CS22892, Barbuy2011_CS31082, Roederer2012_HST, Roederer22}. Similarly, there have been only $\sim15$ RPE stars with a Pt abundance determined \citep{Cowan1996_PtOsPd_UV, Westin2000, Cowan2002, Sneden2003_CS22892, Cowan2005_largeSample, Barbuy2011_CS31082, Roederer2012_HST, Roederer2012_HD160617, Roederer22}. Therefore, the Au and Pt abundances of \thisStarShort\ are highly valuable. Indeed, we find that our results on the abundances of these elements, along with that of Os, uphold the case for the extension of the universal \rproc\ pattern to the third \rproc\ peak and to Au. 
\par
We also find that, with [Au/Fe]$\ = +1.37$, \thisStarShort\ has the highest enhancement in gold discovered in any star thus far. However, we note that this enhancement is not substantially different from what is found in other stars, which have $+0.80 <$ [Au/Fe] $< +1.28$. Such enhancements in Au ([Au/Fe]$\ > 0.0$) remain at odds with the current Galactic chemical-enrichment models, which predict an under-production of [Au/Fe] levels by a factor of 5, or by $0.80$ dex \citep{Kobayashi+2020}. It is possible that such a discrepancy exists due to the observational bias of having Au abundance determined for only a few stars, which are also RPE. Therefore, there is an urgent need for more abundance determination of Au in metal-poor stars, which will require access to high-resolution space-based observations in the UV. To that end, this study serves as an important contribution.

\section{Conclusions}\label{sec:conclusion}
We have performed a detailed chemical-abundance analysis of a very metal-poor ([Fe/H] $=-2.31$) turn-off RPE ([Eu/Fe] $= 1.37$) star, \thisStar. For this, we obtained high-resolution HST/STIS and Magellan/MIKE spectroscopic data covering the UV ($\sim$ $2275$ - $3119$ \AA) and optical ($3350$ - $9500$ \AA) wavelengths. \thisStarShort\ may be the warmest highly RPE star studied in both the UV and optical domains thus far. As a result of the wide wavelength coverage, we determined abundances for 41 elements in total. We have identified several key chemical signatures that contribute to our understanding of the origin of these elements, especially the \rproc\ elements, and motivate future theoretical and observational studies.    
\begin{itemize}
    \item We determined abundances for 16 light elements, including 26 individual species. We found that the light elements follow the typical trends of metal-poor stars (bottom panel of Figure \ref{fig:Fe_groupElements}), indicating that their origin was primarily core-collapse supernovae.
    \item We found that the most significant NLTE corrections are for \ion{Al}{I}, \ion{K}{I},\ion{Cr}{I}, and \ion{Mn}{I} abundances. We show the LTE and NLTE-corrected abundances of individual species in the middle panel of Figure \ref{fig:Fe_groupElements}.
    \item  We found that, even though the NLTE correction for \ion{Al}{I} is large ($+0.47$ dex), it is still not sufficiently large to enable agreement with the LTE abundance of the dominant species, \ion{Al}{II}, which remains $\sim0.4$ dex ($\sim2\sigma$) higher than the NLTE-corrected \ion{Al}{I} abundance. Based on this result, we urge the community to investigate the NLTE effects for \ion{Al}{I} and \ion{Al}{II} further. 
    \item Among the \rproc\ elements, we determined detailed abundances for 23 elements and upper limits for 6 elements, ranging from the first \rproc\ peak to the actinides (Figure \ref{fig:rprocessPattern}). We found that \thisStarShort, like many other RPE stars, exhibits variations in its pattern of light \rproc-elements. Specifically, Ge and Y are found to deviate from the lanthanide-scaled S.S. \rproc\ pattern. However, we find that taking into account NLTE effects for the low-excitation \ion{Y}{II} lines would enable a better agreement with the S.S. \rproc\ pattern. On the other hand, Sr, Y, Zr, and Mo follow the benchmark light \rproc-pattern of metal-poor stars identified by \cite{Roederer2022b} (Figure \ref{fig:zrscaled}), adding evidence for a common source of light \rproc\ elements in the early Universe.  
    \item We found that [Ge/Fe] $= +0.10$ is significantly higher than observed in $\sim20$ metal-poor stars thus far, which exhibit [Ge/Fe] $\sim -0.90$ with a standard deviation of $\sim\ 0.28$ dex. The similar [Ge/Fe] ratios observed in metal-poor stars has indicated a common origin for Ge and Fe in the early Universe, for instance $\alpha$-rich freezeout. Therefore, the high [Ge/Fe] may indicate contribution from a production mechanism decoupled from that of Fe, such as the $\nu p$- or $r$-process. 
    \item Similarly, the Cd abundance observed for this star is the highest to date, adding to the dispersion observed in the Cd abundances of metal-poor stars, which still remains unexplained (Figure \ref{fig:zrscaled}). Here, we found that the \eps(Cd), \eps(Cd/Zr), and \eps(Cd/Eu) abundances of metal-poor stars are correlated to \eTeff\ and \logg\ of the stars, suggesting that the \ion{Cd}{I} line may be severely affected by NLTE effects, causing the observed dispersion (Figure \ref{fig:Cd}). Therefore, we urge the community to theoretically study the NLTE effects for \ion{Cd}{I} and \ion{Cd}{II}, as well as revisit the astrophysical origin of this element. 
    \item We found that the lanthanides generally follow the universal \rproc-pattern. However, the Eu abundance was found to be mildly discrepant from the universal pattern, which we found explainable by NLTE effects.
    \item Additionally, the abundances of Os, Pt, and Au, uphold the case for the extension of the universal \rproc-pattern to the third \rproc\ peak (Figure \ref{fig:rprocessPattern}) -- while this case has been suggested by previous studies, it remains to be firmly established due to the scarce number of abundances of elements like Pt and Au. 
    \item The abundance determination of Au marks \thisStarShort\ as only the 6th star with Au abundance determined. The enhancement in Au, relative to Fe, follows that of other metal-poor stars. These observed enhancements are five times higher than suggested by current Galactic chemical-evolution models, motivating the need for more abundance determinations of Au in metal-poor stars. 
  
\end{itemize}
Overall, as part of the \textit{R}-Process Alliance effort, this study adds to the sparse but growing number of RPE stars with extensive and detailed inventory of chemical abundances. We anticipate more such studies with UV and optical spectroscopy in the near future, especially as an increasing number of \rproc-enhanced stars are identified \citep[e.g.,][]{rpa1, rpa2, RPA3, rpa4}. However, as highlighted in this study, it will also be necessary to advance the theoretical NLTE studies, especially of \rproc\ elements, in concurrence with the astrophysical studies, which critically depend on the reliable abundance determination of these elements. 

\section*{Acknowledgements}
This research is based on observations made with the NASA/ESA Hubble Space Telescope obtained from the Space Telescope Science Institute, which is operated by the Association of Universities for Research in Astronomy, Inc., under NASA contract NAS 5–26555. These observations are associated with program 15951. 
S.P.S acknowledges Jamie Tayar and Guilherme Limberg for helpful conversations.
S.P.S and R.E. acknowledge support from NASA grant GO-15951 from the Space Telescope Science Institute, which is operated by the Association of Universities for Research in Astronomy, Inc., under NASA contract NAS5–26555.
R.E acknowledges support NSF grant AST-2206263.
I.U.R acknowledges NSF grants AST~1815403 and AST~2205847, NASA Astrophysics Data Analysis Program grant 80NSSC21K0627, as well as NASA grants GO-15657, GO-15951, and AR-16630 from the Space Telescope Science Institute, which is operated by the Association of Universities for Research in Astronomy, Incorporated, under NASA contract NAS5-26555. 
TTH acknowledges support from the Swedish Research Council (VR 2021-05556). The work of V.M.P. is supported by NOIRLab, which is managed by the Association of Universities for Research in Astronomy (AURA) under a cooperative agreement with the National Science Foundation.
T.C.B. acknowledges partial support from grant PHY 14-30152; Physics Frontier Center/JINA Center for the Evolution of the Elements (JINA-CEE), and from OISE-1927130: The International Research Network for Nuclear Astrophysics (IReNA), awarded by the US National Science Foundation. 
A.P.J. acknowledges support from NSF grant AST-2206264.
C.M.S acknowledges support from the NSF grant AST 2206379.


\section*{Data Availability}
The UV spectrum of \thisStar\ is publicly available on the Mikulski Archive for Space Telescopes under the proposal ID 15951.



\bibliographystyle{mnras}
\bibliography{main} 




\appendix
\section{Linelist}
\onecolumn
\input{tables/test_latex_wUnc.txt}
\twocolumn


\bsp	
\label{lastpage}
\end{document}

%% file: tables/abund_latex_wUnc.txt
\begin{table*}
\begin{tabular}{ccccccccc}
\hline
\hline
Species & $\log\epsilon$(X)$\mathrm{_{UV}}$ & $\sigma_\mathrm{stat,UV}$ & N$\mathrm{_{UV}}$ & $\log\epsilon$(X)$\mathrm{_{op}}$ & $\sigma_\mathrm{stat,OP}$ & N$\mathrm{_{OP}}$ & $\log\epsilon$(X) & $\Delta$ NLTE$_\mathrm{corr}$ \\
\hline
LiI & -- & -- & -- & 2.37 & 0.20 & 1 & 2.37 & $-$0.018 \\
OI & -- & -- & -- & 7.08 & 0.05 & 3 & 7.08 & $-$0.10 \\
NaI & -- & -- & -- & 3.99 & 0.07 & 4 & 3.99 & $-$0.10 \\
MgI & 5.66 & 0.04 & 2 & 5.73 & 0.09 & 7 & 5.72 & 0.10 \\
MgII & 5.70 & 0.20 & 1 & -- & -- & -- & 5.70 & -- \\
AlI & -- & -- & -- & 3.32 & 0.17 & 2 & 3.32 & 0.47 \\
AlII & 4.18 & 0.20 & 1 & -- & -- & -- & 4.18 & -- \\
SiI & 5.58 & 0.20 & 1 & 5.53 & 0.20 & 1 & 5.56 & 0.003 \\
SiII & 5.60 & 0.01 & 2 & 5.47 & 0.20 & 1 & 5.55 & -- \\
SI & -- & -- & -- & <5.98 & -- & 5 & <5.98 & -- \\
KI & -- & -- & -- & 3.33 & 0.20 & 1 & 3.33 & -0.28 \\
CaI & -- & -- & -- & 4.48 & 0.09 & 24 & 4.48 & 0.14 \\
ScII & 1.13 & 0.20 & 1 & 1.10 & 0.05 & 12 & 1.10 & -- \\
TiI & -- & -- & -- & 3.2 & 0.10 & 12 & 3.2 & 0.14 \\
TiII & 3.10 & 0.10 & 4 & 3.12 & 0.15 & 34 & 3.12 & 0.04 \\
VI & -- & -- & -- & 2.05 & 0.2 & 1 & 2.05 & -- \\
VII & 1.99 & 0.08 & 6 & 1.97 & 0.05 & 7 & 1.98 & -- \\
CrI & 3.22 & 0.03 & 2 & 3.34 & 0.05 & 6 & 3.31 & 0.25 \\
CrII & 3.46 & 0.05 & 11 & 3.31 & 0.02 & 3 & 3.43 & 0.04 \\
MnI & 2.89 & 0.22 & 3 & 2.81 & 0.02 & 5 & 2.84 & 0.28 \\
MnII & 2.97 & 0.14 & 2 & 2.88 & 0.07 & 4 & 2.91 & $-$0.025 \\
FeI & 5.16 & 0.11 & 21 & 5.19 & 0.1 & 156 & 5.19 & 0.17 \\
FeII & 5.10 & 0.12 & 22 & 5.23 & 0.05 & 14 & 5.15 & 0.0 \\
CoI & 2.95 & 0.20 & 1 & 2.87 & 0.12 & 10 & 2.88 & -- \\
CoII & 2.61 & 0.09 & 9 & -- & -- & -- & 2.61 & -- \\
NiI & 3.92 & 0.12 & 3 & 3.94 & 0.16 & 18 & 3.94 & -- \\
NiII & 3.94 & 0.04 & 5  & -- & -- & -- & 3.94 & -- \\
ZnI & -- & -- & -- & 2.32 & 0.02 & 2 & 2.32 & 0.18 \\
GeI & 1.40 & 0.20 & 1 & -- & -- & -- & 1.40 & -- \\
AsI & <0.84 & -- & 1 & -- & -- & -- & <0.84 & -- \\
RbI & -- & -- & -- & <2.42 & -- & 5 & <2.42 & -- \\
SrII & -- & -- & -- & 1.06 & 0.08 & 2 & 1.06 & -- \\
YII & 0.58 & 0.20 & 1 & 0.46 & 0.13 & 9 & 0.48 & -- \\
ZrII & 1.09 & 0.09 & 5 & 1.15 & 0.07 & 10 & 1.13 & -- \\
NbII & <0.93 & -- & 1 & -- & -- & -- & <0.93 & -- \\
MoII & 0.75 & 0.20 & 1 & -- & -- & -- & 0.75 & -- \\
RuII & <1.54 & -- & 1 & -- & -- & -- & <1.54 & -- \\
RhI & -- & -- & -- & <1.52 & -- & 5 & <1.52 & -- \\
PdI & -- & -- & -- & <1.34 & -- & 5 & <1.34 & -- \\
AgI & -- & -- & -- & <1.38 & -- & 5 & <1.38 & -- \\
CdI & 0.63 & 0.20 & 1 & -- & -- & -- & 0.63 & -- \\
InII & <0.83 & -- & 1 & -- & -- & -- & <0.83 & -- \\
SnI & <2.94 & -- & 1 & -- & -- & -- & <2.94 & -- \\
TeI & <1.83 & -- & 1 & -- & -- & -- & <1.83 & -- \\
BaII & -- & -- & -- & 0.48 & 0.06 & 5 & 0.48 & -- \\
LaII & -- & -- & -- & $-$0.16 & 0.06 & 5 & $-$0.16 & -- \\
CeII & -- & -- & -- & 0.24 & 0.03 & 2 & 0.24 & -- \\
PrII & -- & -- & -- & $-$0.40 & 0.20 & 1 & $-$0.40 & -- \\
NdII & -- & -- & -- & 0.17 & 0.12 & 6 & 0.17 & -- \\
SmII & -- & -- & -- & 0.18 & 0.2 & 1 & 0.18 & -- \\
EuII & -- & -- & -- & $-$0.46 & 0.04 & 5 & $-$0.46 & -- \\
GdII & 0.40 & 0.20 & 1 & 0.38 & 0.04 & 2 & 0.38 & -- \\
TbII & -- & -- & -- & <$-$0.06 & -- & 6 & <$-$0.06 & -- \\
DyII & -- & -- & -- & 0.26 & 0.09 & 5 & 0.26 & -- \\
HoII & -- & -- & -- & $-$0.36 & 0.11 & 2 & $-$0.36 & -- \\
ErII & -- & -- & -- & 0.12 & 0.19 & 2 & 0.12 & -- \\
TmII & -- & -- & -- & $-$0.67 & 0.20 & 1 & $-$0.67 & -- \\
YbII & -- & -- & -- & $-$0.27 & 0.20 & 1 & $-$0.27 & -- \\
LuII & -0.70 & 0.20 & 1 & -- & -- & -- & $-$0.70 & -- \\
HfII & -- & -- & -- & <0.67 & -- & 5 & <0.67 & -- \\
OsII & 0.49 & 0.20 & 1 & -- & -- & -- & 0.49 & -- \\
IrI & <1.01 & -- & 1 & -- & -- & -- & <1.01 & -- \\
PtI & 0.60 & 0.13 & 2 & -- & -- & -- & 0.60 & -- \\
AuI & -0.06 & 0.07 & 2 & -- & -- & -- & $-$0.06 & -- \\
ThII & -- & -- & -- & <$-$0.38 & -- & 6 & <$-$0.38 & -- \\
UII & -- & -- & -- & <0.43 & -- & 5 & <0.43 & -- \\
C-H & -- & -- & -- & 6.70 & 0.20 & 1 & 6.70 & -- \\
C-N & -- & -- & -- & <7.73 & -- & 5 & <7.73 & -- \\
\end{tabular}
\caption{\label{tab:UVOP_abund} Mean abundances ($\log\epsilon(X)$), statistical uncertainty ($\sigma_\mathrm{stat}$), and the number of lines used (N) for each species in the UV and optical (op) domains. Also listed are the NLTE corrections for the mean abundances of each species, UV and optical lines considered together.}
\end{table*}

%% file: tables/adopted_abund_latex.txt
\begin{table}
\begin{tabular}{ccccccc}
\hline
\hline
\\
Element & $\log\epsilon$(X)$_\odot$ & $\log\epsilon$(X) & [X/Fe] & $\sigma\mathrm{_{stat}}$ & $\sigma\mathrm{_{sys}}$ & $\sigma\mathrm{_{total}}$ \\
\\
\hline
\\
Li & 1.05 & 2.35 & +3.65 & 0.20 & 0.05 & 0.21 \\
O & 8.69 & 6.98 & +0.64 & 0.05 & 0.05 & 0.07 \\
Na & 6.24 & 3.89 & $-$0.00 & 0.07 & 0.06 & 0.09 \\
Mg & 7.6 & 5.7 & +0.45 & 0.20 & 0.08 & 0.22 \\
Al & 6.45 & 4.18 & +0.08 & 0.20 & 0.06 & 0.21 \\
Si & 7.51 & 5.55 & +0.39 & 0.08 & 0.05 & 0.09 \\
S & 7.12 & <5.98 & -- & -- & -- & -- \\
K & 5.03 & 3.05 & +0.37 & 0.20 & 0.05 & 0.21 \\
Ca & 6.34 & 4.62 & +0.63 & 0.09 & 0.05 & 0.10 \\
Sc & 3.15 & 1.10 & +0.30 & 0.05 & 0.05 & 0.07 \\
Ti & 4.95 & 3.12 & +0.52 & 0.14 & 0.07 & 0.16 \\
V & 3.93 & 1.98 & +0.40 & 0.07 & 0.06 & 0.09 \\
Cr & 5.64 & 3.43 & +0.14 & 0.07 & 0.07 & 0.10 \\
Mn & 5.43 & 2.91 & $-$0.17 & 0.08 & 0.05 & 0.09 \\
Fe & 7.5 & 5.15 & +0.00 & 0.12 & 0.06 & 0.13 \\
Co & 4.99 & 2.61 & $-$0.03 & 0.09 & 0.10 & 0.14 \\
Ni & 6.22 & 3.94 & +0.07 & 0.04 & 0.09 & 0.1 \\
Zn & 4.56 & 2.50 & +0.29 & 0.02 & 0.04 & 0.04 \\
Ge & 3.65 & 1.40 & +0.10 & 0.20 & 0.17 & 0.26 \\
As & 2.3 & <0.84 & -- & -- & -- & -- \\
Rb & 2.52 & <2.42 & -- & -- & -- & -- \\
Sr & 2.87 & 1.06 & +0.54 & 0.08 & 0.18 & 0.20 \\
Y & 2.21 & 0.48 & +0.62 & 0.13 & 0.06 & 0.14 \\
Zr & 2.58 & 1.13 & +0.90 & 0.08 & 0.06 & 0.10 \\
Nb & 1.46 & <0.93 & -- & -- & -- & -- \\
Mo & 1.88 & 0.75 & +1.22 & 0.2 & 0.05 & 0.21 \\
Ru & 1.75 & <1.54 & -- & -- & -- & -- \\
Rh & 0.91 & <1.52 & -- & -- & -- & -- \\
Pd & 1.57 & <1.34 & -- & -- & -- & -- \\
Ag & 0.94 & <1.38 & -- & -- & -- & -- \\
Cd & 1.71 & 0.63 & +1.27 & 0.20 & 0.13 & 0.24 \\
In & 0.8 & <0.83 & -- & -- & -- & -- \\
Sn & 0.8 & <0.83 & -- & -- & -- & -- \\
Te & 2.18 & <1.83 & -- & -- & -- & -- \\
Ba & 2.18 & 0.48 & +0.65 & 0.06 & 0.05 & 0.08 \\
La & 1.1 & -0.16 & +1.09 & 0.06 & 0.06 & 0.09 \\
Ce & 1.58 & 0.24 & +1.01 & 0.03 & 0.05 & 0.05 \\
Pr & 0.72 & -0.40 & +1.23 & 0.20 & 0.20 & 0.28 \\
Nd & 1.42 & 0.17 & +1.10 & 0.12 & 0.11 & 0.16 \\
Sm & 0.96 & 0.18 & +1.57 & 0.20 & 0.07 & 0.21 \\
Eu & 0.52 & -0.46 & +1.37 & 0.04 & 0.05 & 0.07 \\
Gd & 1.07 & 0.38 & +1.66 & 0.03 & 0.15 & 0.15 \\
Tb & 0.30 & <-0.06 & -- & -- & -- & -- \\
Dy & 1.10 & 0.26 & +1.51 & 0.09 & 0.09 & 0.13 \\
Ho & 0.48 & -0.36 & +1.51 & 0.11 & 0.07 & 0.13 \\
Er & 0.92 & 0.12 & +1.55 & 0.19 & 0.10 & 0.21 \\
Tm & 0.10 & -0.67 & +1.58 & 0.20 & 0.06 & 0.21 \\
Yb & 0.84 & -0.27 & +1.24 & 0.20 & 0.05 & 0.21 \\
Lu & 0.10 & -0.7 & +1.55 & 0.20 & 0.17 & 0.26 \\
Hf & 0.85 & <0.67 & -- & -- & -- & -- \\
Os & 1.40 & 0.49 & +1.44 & 0.20 & 0.29 & 0.35 \\
Ir & 1.38 & <1.01 & -- & -- & -- & -- \\
Pt & 1.62 & 0.6 & +1.33 & 0.13 & 0.13 & 0.19 \\
Au & 0.92 & -0.06 & +1.37 & 0.07 & 0.12 & 0.14 \\
Th & 0.02 & <-0.38 & -- & -- & -- & -- \\
U & -0.54 & <0.43 & -- & -- & -- & -- \\
C-H & 8.43 & 6.7 & +0.62 & 0.20 & 0.12 & 0.23 \\
C-N & 7.83 & <7.73 & -- & -- & -- & -- \\
\end{tabular}
\caption{\label{tab:adopted_abund}Adopted and recommended abundances for \thisStar, along with statistical ($\sigma_\mathrm{stat}$), systematic ($\sigma_\mathrm{sys}$), and total uncertainty ($\sigma_\mathrm{tot}$). The Solar abundances of the elements ($\log\epsilon$(X)$_\odot$) are taken from \protect \cite{Asplund2009}.}
\end{table}

%% file: tables/test_latex_wUnc.txt
\begin{longtable}{ccccccc}
\hline
\hline
\\
Species & Wavelength (\AA) & $\chi$ (eV) & $\log\ gf$ & Type & $\log\epsilon$(X) & $\sigma_{sys}$ \\
\\
\hline
LiI & 6707.8 & 0.0 & 0.17 & syn & 2.37 & 0.05 \\
OI & 7771.94 & 9.15 & 0.37 & syn & 7.12 & 0.05 \\
OI & 7774.17 & 9.15 & 0.22 & syn & 7.09 & 0.05 \\
OI & 7775.39 & 9.15 & 0.00 & syn & 7.04 & 0.05 \\
NaI & 5889.95 & 0.0 & 0.11 & eqw & 4.03 & 0.09 \\
NaI & 5895.92 & 0.0 & -0.19 & eqw & 3.92 & 0.07 \\
NaI & 8183.26 & 2.1 & 0.24 & eqw & 3.93 & 0.03 \\
NaI & 8194.81 & 2.1 & 0.54 & eqw & 4.07 & 0.03 \\
MgI & 2736.54 & 2.72 & -1.01 & syn & 5.64 & 0.05 \\
MgI & 2936.74 & 2.71 & -2.27 & syn & 5.68 & 0.06 \\
MgI & 4057.51 & 4.35 & -0.90 & syn & 5.85 & 0.02 \\
MgI & 4167.27 & 4.35 & -0.74 & syn & 5.88 & 0.02 \\
MgI & 4571.1 & 0.0 & -5.62 & syn & 5.72 & 0.07 \\
MgI & 4702.99 & 4.35 & -0.44 & syn & 5.75 & 0.03 \\
MgI & 5172.68 & 2.71 & -0.36 & syn & 5.64 & 0.08 \\
MgI & 5183.60 & 2.72 & -0.17 & syn & 5.63 & 0.09 \\
MgI & 5528.4 & 4.35 & -0.55 & syn & 5.65 & 0.03 \\
MgII & 2928.63 & 4.42 & -0.53 & syn & 5.7 & 0.08 \\
AlI & 3943.0 & 0.0 & -0.64 & syn & 3.42 & 0.07 \\
AlI & 3961.52 & 0.01 & -0.33 & syn & 3.23 & 0.06 \\
AlII & 2669.16 & 0.0 & -4.98 & syn & 4.18 & 0.06 \\
SiI & 2987.64 & 0.78 & -1.97 & syn & 5.58 & 0.08 \\
SiI & 3905.52 & 1.91 & -1.04 & syn & 5.53 & 0.1 \\
SiII & 2334.41 & 0.0 & -5.09 & syn & 5.59 & 0.05 \\
SiII & 2350.17 & 0.04 & -5.12 & syn & 5.6 & 0.07 \\
SiII & 6371.36 & 8.12 & -0.08 & syn & 5.47 & 0.04 \\
SI & 6757.13 & 7.87 & -0.13 & syn & <5.98 & -- \\
KI & 7698.96 & 0.0 & -0.18 & eqw & 3.33 & 0.05 \\
CaI & 4226.74 & 0.0 & 0.24 & eqw & 4.36 & 0.11 \\
CaI & 4283.01 & 1.89 & -0.2 & eqw & 4.4 & 0.05 \\
CaI & 4318.65 & 1.9 & -0.21 & eqw & 4.34 & 0.05 \\
CaI & 4425.44 & 1.88 & -0.41 & eqw & 4.49 & 0.05 \\
CaI & 4434.96 & 1.89 & -0.06 & eqw & 4.43 & 0.05 \\
CaI & 4435.69 & 1.89 & -0.55 & eqw & 4.56 & 0.04 \\
CaI & 4454.78 & 1.9 & 0.26 & eqw & 4.35 & 0.05 \\
CaI & 4455.89 & 1.9 & -0.55 & eqw & 4.46 & 0.05 \\
CaI & 4578.55 & 2.52 & -0.67 & eqw & 4.58 & 0.04 \\
CaI & 5261.71 & 2.52 & -0.6 & eqw & 4.61 & 0.04 \\
CaI & 5512.98 & 2.93 & -0.45 & eqw & 4.59 & 0.03 \\
CaI & 5581.97 & 2.52 & -0.58 & eqw & 4.49 & 0.04 \\
CaI & 5588.76 & 2.52 & 0.3 & eqw & 4.43 & 0.04 \\
CaI & 5590.12 & 2.52 & -0.6 & eqw & 4.58 & 0.04 \\
CaI & 5857.45 & 2.93 & 0.17 & eqw & 4.5 & 0.03 \\
CaI & 6102.72 & 1.88 & -0.81 & eqw & 4.39 & 0.05 \\
CaI & 6122.22 & 1.89 & -0.33 & eqw & 4.46 & 0.05 \\
CaI & 6162.17 & 1.9 & -0.11 & eqw & 4.46 & 0.05 \\
CaI & 6169.06 & 2.52 & -0.87 & eqw & 4.62 & 0.04 \\
CaI & 6169.56 & 2.53 & -0.6 & eqw & 4.6 & 0.04 \\
CaI & 6439.07 & 2.52 & 0.33 & eqw & 4.45 & 0.04 \\
CaI & 6449.81 & 2.52 & -0.55 & eqw & 4.52 & 0.04 \\
CaI & 6499.65 & 2.52 & -0.81 & eqw & 4.31 & 0.04 \\
CaI & 6717.69 & 2.71 & -0.58 & eqw & 4.46 & 0.04 \\
ScII & 2551.0 & -- & -- & syn & 1.13 & 0.04 \\
ScII & 3572.53 & 0.02 & 0.27 & syn & 1.17 & 0.09 \\
ScII & 3576.34 & 0.01 & 0.01 & syn & 1.13 & 0.09 \\
ScII & 3590.47 & 0.02 & -0.55 & syn & 1.1 & 0.04 \\
ScII & 3630.74 & 0.01 & 0.22 & syn & 0.99 & 0.05 \\
ScII & 4246.81 & 0.32 & 0.24 & syn & 1.15 & 0.05 \\
ScII & 4314.08 & 0.62 & -0.1 & syn & 1.14 & 0.02 \\
ScII & 4320.73 & 0.6 & -0.25 & syn & 1.1 & 0.04 \\
ScII & 4324.98 & 0.59 & -0.44 & syn & 1.02 & 0.04 \\
ScII & 4374.45 & 0.62 & -0.42 & syn & 1.12 & 0.03 \\
ScII & 4400.38 & 0.6 & -0.54 & syn & 1.09 & 0.04 \\
ScII & 4415.54 & 0.59 & -0.67 & syn & 1.03 & 0.04 \\
ScII & 4670.4 & 1.36 & -0.58 & syn & 1.15 & 0.04 \\
TiI & 3989.76 & 0.02 & -0.13 & eqw & 3.18 & 0.07 \\
TiI & 3998.64 & 0.05 & 0.02 & eqw & 3.13 & 0.07 \\
TiI & 4512.73 & 0.84 & -0.4 & eqw & 3.29 & 0.06 \\
TiI & 4518.02 & 0.83 & -0.25 & eqw & 3.19 & 0.06 \\
TiI & 4533.24 & 0.85 & 0.54 & eqw & 3.1 & 0.06 \\
TiI & 4534.78 & 0.84 & 0.35 & eqw & 3.1 & 0.06 \\
TiI & 4681.91 & 0.05 & -1.01 & eqw & 3.38 & 0.07 \\
TiI & 4981.73 & 0.84 & 0.57 & eqw & 3.05 & 0.06 \\
TiI & 4991.07 & 0.84 & 0.45 & eqw & 3.13 & 0.06 \\
TiI & 4999.5 & 0.83 & 0.32 & eqw & 3.22 & 0.06 \\
TiI & 5173.74 & 0.0 & -1.06 & eqw & 3.36 & 0.07 \\
TiI & 5192.97 & 0.02 & -0.95 & eqw & 3.31 & 0.07 \\
TiII & 2761.29 & 1.08 & -1.35 & syn & 3.0 & 0.07 \\
TiII & 2841.93 & 0.61 & -0.59 & syn & 3.18 & 0.09 \\
TiII & 3029.73 & 1.57 & -0.35 & syn & 3.04 & 0.04 \\
TiII & 3048.76 & 1.58 & -1.18 & syn & 3.2 & 0.07 \\
TiII & 3340.34 & 0.11 & -0.53 & eqw & 3.26 & 0.18 \\
TiII & 3343.76 & 0.15 & -1.18 & eqw & 3.03 & 0.11 \\
TiII & 3372.79 & 0.01 & 0.28 & eqw & 3.48 & 0.14 \\
TiII & 3387.83 & 0.03 & -0.41 & eqw & 3.46 & 0.19 \\
TiII & 3456.38 & 2.06 & -0.11 & eqw & 2.65 & 0.04 \\
TiII & 3477.18 & 0.12 & -0.95 & eqw & 3.36 & 0.17 \\
TiII & 3489.74 & 0.14 & -2.0 & eqw & 3.21 & 0.07 \\
TiII & 3491.05 & 0.11 & -1.1 & eqw & 3.22 & 0.14 \\
TiII & 3759.29 & 0.61 & 0.28 & eqw & 3.13 & 0.17 \\
TiII & 3761.32 & 0.57 & 0.18 & eqw & 3.06 & 0.17 \\
TiII & 3913.46 & 1.12 & -0.36 & eqw & 3.03 & 0.09 \\
TiII & 4028.34 & 1.89 & -0.92 & eqw & 3.08 & 0.03 \\
TiII & 4394.06 & 1.22 & -1.77 & eqw & 3.1 & 0.04 \\
TiII & 4395.03 & 1.08 & -0.54 & eqw & 3.13 & 0.09 \\
TiII & 4395.84 & 1.24 & -1.93 & eqw & 3.09 & 0.04 \\
TiII & 4399.77 & 1.24 & -1.2 & eqw & 3.06 & 0.04 \\
TiII & 4417.71 & 1.17 & -1.19 & eqw & 3.06 & 0.04 \\
TiII & 4418.33 & 1.24 & -1.99 & eqw & 3.17 & 0.04 \\
TiII & 4443.8 & 1.08 & -0.71 & eqw & 3.09 & 0.07 \\
TiII & 4444.55 & 1.12 & -2.2 & eqw & 3.04 & 0.04 \\
TiII & 4450.48 & 1.08 & -1.52 & eqw & 3.1 & 0.04 \\
TiII & 4464.45 & 1.16 & -1.81 & eqw & 3.08 & 0.04 \\
TiII & 4470.85 & 1.17 & -2.02 & eqw & 2.89 & 0.04 \\
TiII & 4501.27 & 1.12 & -0.77 & eqw & 3.12 & 0.07 \\
TiII & 4533.96 & 1.24 & -0.53 & eqw & 3.06 & 0.07 \\
TiII & 4571.97 & 1.57 & -0.31 & eqw & 3.07 & 0.06 \\
TiII & 4657.2 & 1.24 & -2.29 & eqw & 3.08 & 0.04 \\
TiII & 4708.66 & 1.24 & -2.35 & eqw & 3.16 & 0.04 \\
TiII & 4798.53 & 1.08 & -2.66 & eqw & 3.17 & 0.04 \\
TiII & 5188.69 & 1.58 & -1.05 & eqw & 3.12 & 0.04 \\
TiII & 5226.54 & 1.57 & -1.26 & eqw & 3.07 & 0.04 \\
TiII & 5336.79 & 1.58 & -1.6 & eqw & 3.01 & 0.04 \\
TiII & 5381.02 & 1.57 & -1.97 & eqw & 3.16 & 0.04 \\
TiII & 5418.77 & 1.58 & -2.13 & eqw & 3.32 & 0.04 \\
VI & 4111.78 & 0.3 & 0.4 & syn & 2.05 & 0.1 \\
VII & 2679.32 & 0.03 & -0.63 & syn & 1.96 & 0.05 \\
VII & 2688.71 & 0.04 & -0.98 & syn & 2.02 & 0.09 \\
VII & 2700.93 & 0.04 & -0.37 & syn & 2.05 & 0.05 \\
VII & 2707.86 & 0.0 & -1.39 & syn & 2.11 & 0.13 \\
VII & 2880.03 & 0.35 & -0.64 & syn & 1.9 & 0.1 \\
VII & 2903.07 & 0.32 & -0.7 & syn & 1.89 & 0.05 \\
VII & 3517.3 & 1.13 & -0.24 & syn & 1.89 & 0.03 \\
VII & 3545.2 & 1.09 & -0.32 & syn & 2.05 & 0.02 \\
VII & 3715.46 & 1.57 & -0.22 & syn & 2.0 & 0.04 \\
VII & 3727.34 & 1.69 & -0.23 & syn & 1.95 & 0.04 \\
VII & 3903.25 & 1.48 & -0.91 & syn & 1.98 & 0.08 \\
VII & 4005.7 & 1.82 & -0.45 & syn & 1.92 & 0.04 \\
VII & 4023.38 & 1.8 & -0.61 & syn & 2.03 & 0.03 \\
CrI & 3015.2 & 0.96 & -0.2 & syn & 3.23 & 0.07 \\
CrI & 3021.56 & 1.03 & 0.61 & syn & 3.2 & 0.2 \\
CrI & 3908.76 & 1.0 & -1.05 & syn & 3.42 & 0.06 \\
CrI & 4616.12 & 0.98 & -1.19 & syn & 3.4 & 0.02 \\
CrI & 4646.16 & 1.03 & -0.74 & syn & 3.3 & 0.05 \\
CrI & 4652.16 & 1.0 & -1.04 & syn & 3.31 & 0.04 \\
CrI & 5206.04 & 0.94 & 0.02 & syn & 3.29 & 0.08 \\
CrI & 5345.8 & 1.0 & -0.95 & syn & 3.3 & 0.06 \\
CrII & 2666.01 & 1.51 & -0.08 & syn & 3.44 & 0.05 \\
CrII & 2668.71 & 1.49 & -0.55 & syn & 3.42 & 0.0 \\
CrII & 2671.81 & 1.51 & -0.38 & syn & 3.51 & 0.08 \\
CrII & 2677.16 & 1.55 & 0.35 & syn & 3.51 & 0.07 \\
CrII & 2687.09 & 1.51 & -0.62 & syn & 3.5 & 0.02 \\
CrII & 2751.87 & 1.53 & -0.29 & syn & 3.51 & 0.2 \\
CrII & 2856.76 & 2.43 & -0.59 & syn & 3.44 & 0.12 \\
CrII & 2865.33 & 2.42 & -0.71 & syn & 3.52 & 0.06 \\
CrII & 2867.09 & 2.43 & -0.5 & syn & 3.42 & 0.07 \\
CrII & 2876.24 & 1.51 & -0.87 & syn & 3.39 & 0.05 \\
CrII & 2878.45 & 1.55 & -1.27 & syn & 3.38 & 0.07 \\
CrII & 3408.77 & 2.48 & -0.27 & syn & 3.3 & 0.07 \\
CrII & 4588.2 & 4.07 & -0.65 & syn & 3.31 & 0.02 \\
CrII & 4618.81 & 4.07 & -0.89 & syn & 3.33 & 0.03 \\
MnI & 2298.84 & 2.89 & -1.75 & syn & 3.12 & 0.01 \\
MnI & 2610.51 & 3.07 & -0.28 & syn & 2.8 & 0.11 \\
MnI & 2706.14 & 2.95 & -1.34 & syn & 2.75 & 0.06 \\
MnI & 4041.35 & 2.11 & 0.28 & syn & 2.78 & 0.03 \\
MnI & 4055.54 & 2.14 & -0.08 & syn & 2.82 & 0.11 \\
MnI & 4754.04 & 2.28 & -0.08 & syn & 2.8 & 0.06 \\
MnI & 4762.37 & 2.89 & 0.3 & syn & 2.83 & 0.04 \\
MnI & 4783.43 & 2.3 & 0.04 & syn & 2.83 & 0.04 \\
MnII & 2939.31 & 1.17 & 0.11 & syn & 3.05 & 0.03 \\
MnII & 2949.20 & 1.18 & 0.25 & syn & 2.89 & 0.06 \\
MnII & 3441.99 & 1.78 & -0.35 & syn & 2.87 & 0.03 \\
MnII & 3460.32 & 1.81 & -0.63 & syn & 2.87 & 0.07 \\
MnII & 3482.90 & 1.83 & -0.84 & syn & 2.97 & 0.07 \\
MnII & 3488.68 & 1.85 & -0.94 & syn & 2.82 & 0.05 \\
FeI & 2283.66 & 0.11 & -2.22 & syn & 5.23 & 0.1 \\
FeI & 2294.41 & 0.11 & -1.54 & syn & 5.25 & 0.04 \\
FeI & 2485.99 & 0.92 & -1.61 & syn & 5.14 & 0.1 \\
FeI & 2539.36 & 0.92 & -1.79 & syn & 5.2 & 0.1 \\
FeI & 2636.48 & 0.92 & -2.04 & syn & 5.3 & 0.08 \\
FeI & 2641.64 & 0.92 & -1.32 & syn & 4.95 & 0.05 \\
FeI & 2644.0 & 1.01 & -0.91 & syn & 4.94 & 0.05 \\
FeI & 2645.42 & 0.11 & -2.75 & syn & 5.02 & 0.05 \\
FeI & 2647.56 & 0.05 & -2.42 & syn & 5.25 & 0.07 \\
FeI & 2666.4 & 0.96 & -1.87 & syn & 5.2 & 0.07 \\
FeI & 2680.45 & 0.99 & -1.74 & syn & 4.95 & 0.05 \\
FeI & 2690.07 & 0.0 & -2.72 & syn & 5.12 & 0.04 \\
FeI & 2728.02 & 0.92 & -1.46 & syn & 5.16 & 0.09 \\
FeI & 2730.98 & 1.01 & -1.73 & syn & 5.18 & 0.11 \\
FeI & 2838.12 & 0.99 & -1.11 & syn & 5.18 & 0.06 \\
FeI & 2877.3 & 1.48 & -1.3 & syn & 5.16 & 0.03 \\
FeI & 2960.66 & 2.95 & -1.0 & syn & 5.2 & 0.09 \\
FeI & 2966.9 & 0.0 & -0.4 & syn & 5.31 & 0.11 \\
FeI & 3009.57 & 0.92 & -0.76 & syn & 5.25 & 0.1 \\
FeI & 3024.03 & 0.11 & -1.48 & syn & 5.23 & 0.16 \\
FeI & 3059.09 & 0.05 & -0.69 & syn & 5.1 & 0.06 \\
FeI & 3406.8 & 2.22 & -0.96 & eqw & 5.21 & 0.08 \\
FeI & 3440.61 & 0.0 & -0.67 & eqw & 5.27 & 0.19 \\
FeI & 3440.99 & 0.05 & -0.96 & eqw & 5.43 & 0.2 \\
FeI & 3451.91 & 2.22 & -1.0 & eqw & 5.17 & 0.07 \\
FeI & 3490.57 & 0.05 & -1.1 & eqw & 5.36 & 0.2 \\
FeI & 3540.12 & 2.87 & -0.71 & eqw & 5.14 & 0.07 \\
FeI & 3565.38 & 0.96 & -0.13 & eqw & 5.38 & 0.18 \\
FeI & 3608.86 & 1.01 & -0.09 & eqw & 5.35 & 0.18 \\
FeI & 3618.77 & 0.99 & -0.0 & eqw & 5.22 & 0.18 \\
FeI & 3647.84 & 0.92 & -0.14 & eqw & 4.93 & 0.16 \\
FeI & 3727.62 & 0.96 & -0.61 & eqw & 5.08 & 0.15 \\
FeI & 3742.62 & 2.94 & -0.81 & eqw & 5.06 & 0.05 \\
FeI & 3758.23 & 0.96 & -0.01 & eqw & 5.05 & 0.15 \\
FeI & 3786.68 & 1.01 & -2.18 & eqw & 5.18 & 0.07 \\
FeI & 3787.88 & 1.01 & -0.84 & eqw & 5.12 & 0.13 \\
FeI & 3815.84 & 1.48 & 0.24 & eqw & 5.08 & 0.15 \\
FeI & 3849.97 & 1.01 & -0.86 & eqw & 5.21 & 0.14 \\
FeI & 3856.37 & 0.05 & -1.28 & eqw & 5.27 & 0.18 \\
FeI & 3865.52 & 1.01 & -0.95 & eqw & 5.17 & 0.13 \\
FeI & 3878.02 & 0.96 & -0.9 & eqw & 5.22 & 0.14 \\
FeI & 3899.71 & 0.09 & -1.52 & eqw & 5.25 & 0.16 \\
FeI & 3902.95 & 1.56 & -0.44 & eqw & 5.07 & 0.12 \\
FeI & 3920.26 & 0.12 & -1.73 & eqw & 5.32 & 0.15 \\
FeI & 3922.91 & 0.05 & -1.63 & eqw & 5.35 & 0.17 \\
FeI & 3949.95 & 2.18 & -1.25 & eqw & 5.22 & 0.06 \\
FeI & 3977.74 & 2.2 & -1.12 & eqw & 5.12 & 0.06 \\
FeI & 4005.24 & 1.56 & -0.58 & eqw & 5.12 & 0.11 \\
FeI & 4045.81 & 1.49 & 0.28 & eqw & 5.18 & 0.14 \\
FeI & 4058.22 & 3.21 & -1.18 & eqw & 5.39 & 0.05 \\
FeI & 4063.59 & 1.56 & 0.06 & eqw & 5.21 & 0.15 \\
FeI & 4067.98 & 3.21 & -0.53 & eqw & 5.28 & 0.05 \\
FeI & 4070.77 & 3.24 & -0.87 & eqw & 5.36 & 0.05 \\
FeI & 4071.74 & 1.61 & -0.01 & eqw & 5.16 & 0.14 \\
FeI & 4114.44 & 2.83 & -1.3 & eqw & 4.97 & 0.05 \\
FeI & 4132.06 & 1.61 & -0.68 & eqw & 5.26 & 0.11 \\
FeI & 4134.68 & 2.83 & -0.65 & eqw & 5.21 & 0.05 \\
FeI & 4143.87 & 1.56 & -0.51 & eqw & 5.16 & 0.12 \\
FeI & 4147.67 & 1.49 & -2.07 & eqw & 5.26 & 0.07 \\
FeI & 4154.5 & 2.83 & -0.69 & eqw & 5.13 & 0.05 \\
FeI & 4156.8 & 2.83 & -0.81 & eqw & 5.21 & 0.05 \\
FeI & 4157.78 & 3.42 & -0.4 & eqw & 5.14 & 0.05 \\
FeI & 4174.91 & 0.91 & -2.94 & eqw & 5.17 & 0.07 \\
FeI & 4175.64 & 2.85 & -0.83 & eqw & 5.28 & 0.05 \\
FeI & 4181.76 & 2.83 & -0.37 & eqw & 5.19 & 0.06 \\
FeI & 4182.38 & 3.02 & -1.18 & eqw & 5.08 & 0.05 \\
FeI & 4184.89 & 2.83 & -0.87 & eqw & 5.14 & 0.05 \\
FeI & 4187.04 & 2.45 & -0.56 & eqw & 5.15 & 0.06 \\
FeI & 4187.8 & 2.42 & -0.51 & eqw & 5.14 & 0.06 \\
FeI & 4191.43 & 2.47 & -0.67 & eqw & 5.17 & 0.06 \\
FeI & 4195.33 & 3.33 & -0.49 & eqw & 5.21 & 0.05 \\
FeI & 4199.1 & 3.05 & 0.16 & eqw & 5.07 & 0.06 \\
FeI & 4202.03 & 1.49 & -0.69 & eqw & 5.18 & 0.11 \\
FeI & 4216.18 & 0.0 & -3.36 & eqw & 5.36 & 0.09 \\
FeI & 4217.55 & 3.43 & -0.48 & eqw & 5.22 & 0.05 \\
FeI & 4222.21 & 2.45 & -0.91 & eqw & 5.09 & 0.06 \\
FeI & 4227.43 & 3.33 & 0.27 & eqw & 5.19 & 0.05 \\
FeI & 4233.6 & 2.48 & -0.6 & eqw & 5.17 & 0.06 \\
FeI & 4238.81 & 3.4 & -0.23 & eqw & 5.2 & 0.05 \\
FeI & 4250.12 & 2.47 & -0.38 & eqw & 5.14 & 0.06 \\
FeI & 4250.79 & 1.56 & -0.71 & eqw & 5.19 & 0.1 \\
FeI & 4260.47 & 2.4 & 0.08 & eqw & 5.11 & 0.09 \\
FeI & 4271.15 & 2.45 & -0.34 & eqw & 5.13 & 0.07 \\
FeI & 4271.76 & 1.49 & -0.17 & eqw & 5.22 & 0.15 \\
FeI & 4282.4 & 2.18 & -0.78 & eqw & 5.12 & 0.06 \\
FeI & 4325.76 & 1.61 & 0.01 & eqw & 5.11 & 0.14 \\
FeI & 4352.73 & 2.22 & -1.29 & eqw & 5.22 & 0.06 \\
FeI & 4375.93 & 0.0 & -3.0 & eqw & 5.23 & 0.09 \\
FeI & 4388.41 & 3.6 & -0.68 & eqw & 5.21 & 0.04 \\
FeI & 4404.75 & 1.56 & -0.15 & eqw & 5.21 & 0.14 \\
FeI & 4415.12 & 1.61 & -0.62 & eqw & 5.19 & 0.11 \\
FeI & 4427.31 & 0.05 & -2.92 & eqw & 5.25 & 0.09 \\
FeI & 4430.61 & 2.22 & -1.73 & eqw & 5.16 & 0.06 \\
FeI & 4442.34 & 2.2 & -1.23 & eqw & 5.24 & 0.06 \\
FeI & 4443.19 & 2.86 & -1.04 & eqw & 5.08 & 0.05 \\
FeI & 4447.72 & 2.22 & -1.36 & eqw & 5.23 & 0.06 \\
FeI & 4461.65 & 0.09 & -3.19 & eqw & 5.28 & 0.08 \\
FeI & 4466.55 & 2.83 & -0.6 & eqw & 5.15 & 0.05 \\
FeI & 4484.22 & 3.6 & -0.64 & eqw & 5.04 & 0.04 \\
FeI & 4494.56 & 2.2 & -1.14 & eqw & 5.21 & 0.06 \\
FeI & 4531.15 & 1.48 & -2.1 & eqw & 5.19 & 0.07 \\
FeI & 4592.65 & 1.56 & -2.46 & eqw & 5.36 & 0.07 \\
FeI & 4602.94 & 1.49 & -2.21 & eqw & 5.28 & 0.07 \\
FeI & 4619.29 & 3.6 & -1.06 & eqw & 5.07 & 0.04 \\
FeI & 4637.5 & 3.28 & -1.29 & eqw & 5.12 & 0.05 \\
FeI & 4647.43 & 2.95 & -1.35 & eqw & 5.21 & 0.05 \\
FeI & 4668.13 & 3.27 & -1.08 & eqw & 5.17 & 0.05 \\
FeI & 4733.59 & 1.49 & -2.99 & eqw & 5.37 & 0.07 \\
FeI & 4736.77 & 3.21 & -0.67 & eqw & 5.14 & 0.05 \\
FeI & 4871.32 & 2.87 & -0.34 & eqw & 5.07 & 0.05 \\
FeI & 4872.14 & 2.88 & -0.57 & eqw & 5.05 & 0.05 \\
FeI & 4890.76 & 2.88 & -0.38 & eqw & 5.12 & 0.05 \\
FeI & 4891.49 & 2.85 & -0.11 & eqw & 5.11 & 0.06 \\
FeI & 4903.31 & 2.88 & -0.89 & eqw & 5.11 & 0.05 \\
FeI & 4918.99 & 2.86 & -0.34 & eqw & 5.14 & 0.06 \\
FeI & 4920.5 & 2.83 & 0.07 & eqw & 5.11 & 0.07 \\
FeI & 4938.81 & 2.88 & -1.08 & eqw & 5.17 & 0.05 \\
FeI & 4939.69 & 0.86 & -3.25 & eqw & 5.29 & 0.07 \\
FeI & 4946.39 & 3.37 & -1.11 & eqw & 5.09 & 0.05 \\
FeI & 4966.09 & 3.33 & -0.79 & eqw & 5.18 & 0.05 \\
FeI & 5001.86 & 3.88 & -0.01 & eqw & 5.29 & 0.04 \\
FeI & 5006.12 & 2.83 & -0.62 & eqw & 5.1 & 0.05 \\
FeI & 5014.94 & 3.94 & -0.18 & eqw & 5.22 & 0.04 \\
FeI & 5049.82 & 2.28 & -1.36 & eqw & 5.25 & 0.06 \\
FeI & 5051.63 & 0.92 & -2.76 & eqw & 5.33 & 0.07 \\
FeI & 5068.77 & 2.94 & -1.04 & eqw & 5.15 & 0.05 \\
FeI & 5079.74 & 0.99 & -3.24 & eqw & 5.32 & 0.07 \\
FeI & 5083.34 & 0.96 & -2.84 & eqw & 5.13 & 0.07 \\
FeI & 5123.72 & 1.01 & -3.06 & eqw & 5.39 & 0.07 \\
FeI & 5127.36 & 0.92 & -3.25 & eqw & 5.29 & 0.07 \\
FeI & 5133.69 & 4.18 & 0.36 & eqw & 5.13 & 0.04 \\
FeI & 5150.84 & 0.99 & -3.04 & eqw & 5.23 & 0.07 \\
FeI & 5171.6 & 1.49 & -1.72 & eqw & 5.23 & 0.07 \\
FeI & 5191.45 & 3.04 & -0.55 & eqw & 5.19 & 0.05 \\
FeI & 5192.34 & 3.0 & -0.42 & eqw & 5.18 & 0.05 \\
FeI & 5194.94 & 1.56 & -2.02 & eqw & 5.18 & 0.07 \\
FeI & 5215.18 & 3.27 & -0.86 & eqw & 5.06 & 0.05 \\
FeI & 5216.27 & 1.61 & -2.08 & eqw & 5.24 & 0.07 \\
FeI & 5232.94 & 2.94 & -0.06 & eqw & 5.08 & 0.06 \\
FeI & 5266.56 & 3.0 & -0.38 & eqw & 5.15 & 0.05 \\
FeI & 5281.79 & 3.04 & -0.83 & eqw & 5.01 & 0.05 \\
FeI & 5283.62 & 3.24 & -0.45 & eqw & 5.18 & 0.05 \\
FeI & 5324.18 & 3.21 & -0.11 & eqw & 5.11 & 0.05 \\
FeI & 5339.93 & 3.27 & -0.63 & eqw & 4.95 & 0.05 \\
FeI & 5341.02 & 1.61 & -1.95 & eqw & 5.33 & 0.07 \\
FeI & 5364.87 & 4.45 & 0.23 & eqw & 5.16 & 0.04 \\
FeI & 5367.47 & 4.42 & 0.44 & eqw & 5.06 & 0.04 \\
FeI & 5369.96 & 4.37 & 0.54 & eqw & 5.03 & 0.04 \\
FeI & 5371.49 & 0.96 & -1.64 & eqw & 5.3 & 0.1 \\
FeI & 5383.37 & 4.31 & 0.64 & eqw & 5.1 & 0.04 \\
FeI & 5397.13 & 0.92 & -1.98 & eqw & 5.31 & 0.08 \\
FeI & 5405.77 & 0.99 & -1.85 & eqw & 5.32 & 0.08 \\
FeI & 5410.91 & 4.47 & 0.4 & eqw & 5.08 & 0.04 \\
FeI & 5415.2 & 4.39 & 0.64 & eqw & 5.12 & 0.04 \\
FeI & 5429.7 & 0.96 & -1.88 & eqw & 5.34 & 0.08 \\
FeI & 5434.52 & 1.01 & -2.13 & eqw & 5.34 & 0.08 \\
FeI & 5497.52 & 1.01 & -2.82 & eqw & 5.25 & 0.07 \\
FeI & 5501.47 & 0.96 & -3.05 & eqw & 5.38 & 0.07 \\
FeI & 5506.78 & 0.99 & -2.79 & eqw & 5.33 & 0.07 \\
FeI & 5569.62 & 3.42 & -0.52 & eqw & 5.12 & 0.05 \\
FeI & 5586.76 & 3.37 & -0.11 & eqw & 5.08 & 0.05 \\
FeI & 5624.54 & 3.42 & -0.76 & eqw & 5.08 & 0.05 \\
FeI & 5662.52 & 4.18 & -0.41 & eqw & 5.04 & 0.04 \\
FeI & 6003.01 & 3.88 & -1.1 & eqw & 5.35 & 0.04 \\
FeI & 6136.61 & 2.45 & -1.41 & eqw & 5.28 & 0.06 \\
FeI & 6137.69 & 2.59 & -1.35 & eqw & 5.14 & 0.06 \\
FeI & 6191.56 & 2.43 & -1.42 & eqw & 5.15 & 0.06 \\
FeI & 6230.72 & 2.56 & -1.28 & eqw & 5.27 & 0.06 \\
FeI & 6246.32 & 3.6 & -0.77 & eqw & 5.2 & 0.05 \\
FeI & 6252.56 & 2.4 & -1.77 & eqw & 5.24 & 0.06 \\
FeI & 6393.6 & 2.43 & -1.58 & eqw & 5.28 & 0.06 \\
FeI & 6411.65 & 3.65 & -0.59 & eqw & 5.2 & 0.05 \\
FeI & 6421.35 & 2.28 & -2.01 & eqw & 5.34 & 0.06 \\
FeI & 6430.85 & 2.18 & -1.95 & eqw & 5.28 & 0.06 \\
FeI & 6494.98 & 2.4 & -1.24 & eqw & 5.23 & 0.06 \\
FeI & 6677.99 & 2.69 & -1.42 & eqw & 5.31 & 0.06 \\
FeI & 7495.07 & 4.22 & -0.1 & eqw & 5.23 & 0.04 \\
FeII & 2424.39 & 2.58 & -0.94 & syn & 5.2 & 0.04 \\
FeII & 2428.8 & 3.89 & -0.31 & syn & 5.0 & 0.07 \\
FeII & 2434.24 & 5.29 & 0.25 & syn & 5.0 & 0.08 \\
FeII & 2437.65 & 5.2 & -0.36 & syn & 5.1 & 0.08 \\
FeII & 2439.3 & 3.15 & 0.45 & syn & 4.88 & 0.08 \\
FeII & 2444.51 & 2.58 & 0.3 & syn & 5.2 & 0.06 \\
FeII & 2445.57 & 2.7 & 0.05 & syn & 4.84 & 0.08 \\
FeII & 2458.97 & 3.89 & -0.04 & syn & 4.9 & 0.08 \\
FeII & 2461.28 & 3.23 & 0.23 & syn & 5.12 & 0.08 \\
FeII & 2461.86 & 3.22 & 0.34 & syn & 4.95 & 0.08 \\
FeII & 2463.28 & 3.15 & -0.19 & syn & 5.11 & 0.1 \\
FeII & 2465.91 & 3.22 & -0.05 & syn & 5.1 & 0.08 \\
FeII & 2472.61 & 5.55 & 0.47 & syn & 5.2 & 0.08 \\
FeII & 2503.87 & 3.77 & 0.32 & syn & 5.19 & 0.12 \\
FeII & 2572.97 & 2.89 & -1.2 & syn & 5.3 & 0.09 \\
FeII & 2587.95 & 4.15 & 0.23 & syn & 5.15 & 0.07 \\
FeII & 2664.66 & 3.39 & 0.31 & syn & 5.15 & 0.08 \\
FeII & 2718.64 & 6.22 & 0.02 & syn & 5.15 & 0.06 \\
FeII & 2892.83 & 1.08 & -2.7 & syn & 5.2 & 0.03 \\
FeII & 2944.39 & 1.7 & -0.85 & syn & 5.24 & 0.09 \\
FeII & 2984.82 & 1.67 & -0.45 & syn & 5.15 & 0.08 \\
FeII & 2985.55 & 1.72 & -0.89 & syn & 5.12 & 0.07 \\
FeII & 4173.45 & 2.58 & -2.38 & eqw & 5.28 & 0.03 \\
FeII & 4178.86 & 2.58 & -2.51 & eqw & 5.2 & 0.03 \\
FeII & 4233.16 & 2.58 & -2.02 & eqw & 5.26 & 0.04 \\
FeII & 4385.38 & 2.78 & -2.64 & eqw & 5.2 & 0.03 \\
FeII & 4416.82 & 2.78 & -2.57 & eqw & 5.14 & 0.03 \\
FeII & 4491.41 & 2.86 & -2.71 & eqw & 5.17 & 0.03 \\
FeII & 4508.28 & 2.86 & -2.42 & eqw & 5.29 & 0.03 \\
FeII & 4515.34 & 2.84 & -2.6 & eqw & 5.35 & 0.03 \\
FeII & 4555.89 & 2.83 & -2.4 & eqw & 5.23 & 0.03 \\
FeII & 4576.34 & 2.84 & -2.95 & eqw & 5.22 & 0.03 \\
FeII & 4583.83 & 2.81 & -1.94 & eqw & 5.26 & 0.03 \\
FeII & 5197.58 & 3.23 & -2.22 & eqw & 5.23 & 0.03 \\
FeII & 5234.63 & 3.22 & -2.18 & eqw & 5.19 & 0.03 \\
FeII & 5276.0 & 3.2 & -2.01 & eqw & 5.2 & 0.03 \\
CoI & 2316.85 & 0.17 & -1.15 & syn & 2.95 & 0.14 \\
CoI & 3405.12 & 0.43 & 0.29 & syn & 2.98 & 0.09 \\
CoI & 3409.18 & 0.51 & -0.22 & syn & 2.76 & 0.1 \\
CoI & 3433.04 & 0.63 & -0.18 & syn & 2.84 & 0.08 \\
CoI & 3449.17 & 0.58 & -0.12 & syn & 2.75 & 0.1 \\
CoI & 3449.44 & 0.43 & -0.48 & syn & 2.87 & 0.09 \\
CoI & 3489.4 & 0.92 & 0.18 & syn & 2.64 & 0.05 \\
CoI & 3513.48 & 0.1 & -0.79 & syn & 3.09 & 0.07 \\
CoI & 3529.03 & 0.17 & -0.89 & syn & 2.96 & 0.08 \\
CoI & 3995.31 & 0.92 & -0.18 & syn & 2.87 & 0.05 \\
CoI & 4121.32 & 0.92 & -0.33 & syn & 2.94 & 0.06 \\
CoII & 2311.6 & 0.56 & 0.32 & syn & 2.55 & 0.1 \\
CoII & 2326.14 & 0.57 & -0.42 & syn & 2.7 & 0.03 \\
CoII & 2330.36 & 0.61 & -0.51 & syn & 2.8 & 0.06 \\
CoII & 2361.52 & 0.64 & -1.16 & syn & 2.65 & 0.13 \\
CoII & 2393.91 & 0.57 & -0.37 & syn & 2.6 & 0.12 \\
CoII & 2414.07 & 0.57 & -0.37 & syn & 2.6 & 0.06 \\
CoII & 2417.66 & 0.5 & -0.25 & syn & 2.49 & 0.09 \\
CoII & 2464.2 & 1.22 & -0.4 & syn & 2.57 & 0.35 \\
CoII & 2564.0 & 1.33 & 0.03 & syn & 2.55 & 0.0 \\
NiI & 2441.82 & 0.21 & -1.51 & syn & 3.95 & 0.07 \\
NiI & 2984.13 & 0.0 & -1.5 & syn & 3.8 & 0.13 \\
NiI & 2992.59 & 0.02 & -1.22 & syn & 4.0 & 0.1 \\
NiI & 3423.71 & 0.21 & -0.71 & eqw & 4.0 & 0.16 \\
NiI & 3433.56 & 0.03 & -0.67 & eqw & 3.96 & 0.17 \\
NiI & 3437.28 & 0.0 & -1.2 & eqw & 3.93 & 0.13 \\
NiI & 3452.89 & 0.11 & -0.9 & eqw & 4.34 & 0.18 \\
NiI & 3472.54 & 0.11 & -0.79 & eqw & 3.8 & 0.14 \\
NiI & 3483.78 & 0.27 & -1.11 & eqw & 3.75 & 0.1 \\
NiI & 3492.96 & 0.11 & -0.24 & eqw & 4.0 & 0.19 \\
NiI & 3500.85 & 0.17 & -1.27 & eqw & 4.1 & 0.12 \\
NiI & 3524.54 & 0.03 & 0.01 & eqw & 4.01 & 0.19 \\
NiI & 3566.37 & 0.42 & -0.24 & eqw & 3.83 & 0.17 \\
NiI & 3597.7 & 0.21 & -1.1 & eqw & 4.08 & 0.14 \\
NiI & 3783.53 & 0.42 & -1.4 & eqw & 3.97 & 0.08 \\
NiI & 3807.14 & 0.42 & -1.23 & eqw & 3.92 & 0.09 \\
NiI & 4604.99 & 3.48 & -0.24 & eqw & 4.02 & 0.04 \\
NiI & 4648.65 & 3.42 & -0.09 & eqw & 3.88 & 0.04 \\
NiI & 4714.42 & 3.38 & 0.25 & eqw & 3.94 & 0.04 \\
NiI & 5080.53 & 3.65 & 0.32 & eqw & 3.56 & 0.04 \\
NiI & 5476.9 & 1.83 & -0.78 & eqw & 3.9 & 0.06 \\
NiII & 2297.49 & 1.32 & -0.33 & syn & 3.98 & 0.18 \\
NiII & 2350.85 & 1.68 & -2.28 & syn & 3.92 & 0.17 \\
NiII & 2356.40 & 1.86 & -0.83 & syn & 3.91 & 0.07 \\
NiII & 2415.0 & 1.86 & 0.13 & syn & 4.0 & 0.0 \\
NiII & 2437.0 & 1.68 & -0.33 & syn & 3.91 & 0.04 \\
ZnI & 4722.15 & 4.03 & -0.37 & syn & 2.31 & 0.04 \\
ZnI & 4810.53 & 4.08 & -0.15 & syn & 2.33 & 0.04 \\
GeI & 3039.07 & 0.88 & -0.07 & syn & 1.4 & 0.17 \\
AsI & 2288.11 & 1.35 & -0.06 & syn & <0.84 & -- \\
RbI & 7947.6 & 0.0 & -0.16 & syn & <2.42 & -- \\
SrII & 4077.71 & 0.0 & 0.15 & syn & 1.01 & 0.18 \\
SrII & 4215.52 & 0.0 & -0.17 & syn & 1.1 & 0.18 \\
YII & 2422.19 & 0.41 & -0.08 & syn & 0.58 & 0.06 \\
YII & 3549.0 & 0.13 & -0.29 & syn & 0.47 & 0.06 \\
YII & 3600.73 & 0.18 & 0.34 & syn & 0.6 & 0.08 \\
YII & 3611.04 & 0.13 & 0.05 & syn & 0.28 & 0.05 \\
YII & 3950.35 & 0.1 & -0.73 & syn & 0.5 & 0.05 \\
YII & 3982.59 & 0.13 & -0.56 & syn & 0.43 & 0.06 \\
YII & 4235.73 & 0.13 & -1.27 & syn & 0.65 & 0.03 \\
YII & 4398.01 & 0.13 & -0.75 & syn & 0.45 & 0.07 \\
YII & 4883.68 & 1.08 & 0.19 & syn & 0.24 & 0.04 \\
YII & 5205.72 & 1.03 & -0.28 & syn & 0.55 & 0.06 \\
ZrII & 2567.64 & 0.0 & -0.17 & syn & 0.99 & 0.04 \\
ZrII & 2700.13 & 0.1 & -0.08 & syn & 1.05 & 0.0 \\
ZrII & 2745.85 & 0.10 & -0.31 & syn & 1.19 & 0.02 \\
ZrII & 2758.81 & 0.0 & -0.56 & syn & 1.12 & 0.22 \\
ZrII & 2915.99 & 0.47 & -0.5 & syn & 1.12 & 0.1 \\
ZrII & 3430.53 & 0.47 & -0.16 & syn & 1.15 & 0.0 \\
ZrII & 3505.67 & 0.16 & -0.39 & syn & 1.2 & 0.05 \\
ZrII & 3551.95 & 0.1 & -0.36 & syn & 1.24 & 0.07 \\
ZrII & 3998.96 & 0.56 & -0.52 & syn & 1.1 & 0.05 \\
ZrII & 4149.2 & 0.8 & -0.04 & syn & 1.07 & 0.05 \\
ZrII & 4156.23 & 0.71 & -0.78 & syn & 1.0 & 0.06 \\
ZrII & 4161.2 & 0.71 & -0.59 & syn & 1.2 & 0.06 \\
ZrII & 4208.98 & 0.71 & -0.51 & syn & 1.15 & 0.05 \\
ZrII & 4359.73 & 1.24 & -0.51 & syn & 1.25 & 0.09 \\
ZrII & 4496.96 & 0.71 & -0.89 & syn & 1.19 & 0.05 \\
NbII & 2927.81 & 0.51 & 0.16 & syn & <0.93 & -- \\
MoII & 2871.51 & 1.54 & 0.06 & syn & 0.75 & 0.05 \\
RuII & 2456.0 & 1.35 & 0.06 & syn & <1.54 & -- \\
RhI & 3434.89 & 0.0 & 0.44 & syn & <1.52 & -- \\
PdI & 3404.58 & 0.81 & 0.32 & syn & <1.34 & -- \\
AgI & 3382.89 & 0.0 & -0.33 & syn & <1.38 & -- \\
CdI & 2288.02 & 0.0 & 0.11 & syn & 0.63 & 0.13 \\
InII & 2306.06 & 0.0 & -2.3 & syn & <0.83 & -- \\
SnI & 2286.68 & 0.42 & -0.94 & syn & <2.94 & -- \\
TeI & 2385.79 & 0.59 & -0.81 & syn & <1.83 & -- \\
BaII & 4554.03 & 0.0 & 0.17 & syn & 0.55 & 0.07 \\
BaII & 4934.08 & 0.0 & -0.16 & syn & 0.4 & 0.07 \\
BaII & 5853.68 & 0.6 & -0.91 & syn & 0.47 & 0.03 \\
BaII & 6141.71 & 0.7 & -0.08 & syn & 0.5 & 0.06 \\
BaII & 6496.9 & 0.6 & -0.38 & syn & 0.47 & 0.04 \\
LaII & 3949.1 & 0.4 & 0.49 & syn & -0.2 & 0.06 \\
LaII & 3988.51 & 0.4 & 0.21 & syn & -0.2 & 0.06 \\
LaII & 4077.34 & 0.24 & -0.06 & syn & -0.05 & 0.07 \\
LaII & 4086.71 & 0.0 & -0.07 & syn & -0.18 & 0.05 \\
LaII & 4123.22 & 0.32 & 0.13 & syn & -0.16 & 0.06 \\
CeII & 4460.21 & 0.48 & 0.28 & syn & 0.22 & 0.02 \\
CeII & 4562.36 & 0.48 & 0.21 & syn & 0.25 & 0.07 \\
PrII & 4225.32 & 0.0 & 0.32 & syn & -0.4 & 0.2 \\
NdII & 3900.22 & 0.47 & 0.1 & syn & 0.3 & 0.12 \\
NdII & 4012.24 & 0.63 & 0.81 & syn & 0.07 & 0.19 \\
NdII & 4109.45 & 0.32 & 0.35 & syn & 0.03 & 0.07 \\
NdII & 4156.08 & 0.18 & 0.16 & syn & 0.1 & 0.07 \\
NdII & 4177.32 & 0.06 & -0.1 & syn & 0.34 & 0.09 \\
NdII & 4303.57 & 0.0 & 0.08 & syn & 0.17 & 0.13 \\
SmII & 4329.02 & 0.18 & -0.51 & syn & 0.18 & 0.07 \\
EuII & 3819.67 & 0.0 & 0.51 & syn & -0.51 & 0.04 \\
EuII & 3907.11 & 0.21 & 0.17 & syn & -0.41 & 0.04 \\
EuII & 4129.72 & 0.0 & 0.22 & syn & -0.47 & 0.03 \\
EuII & 4205.04 & 0.0 & 0.21 & syn & -0.44 & 0.04 \\
EuII & 4435.58 & 0.21 & -0.11 & syn & -0.45 & 0.11 \\
GdII & 3032.84 & 0.08 & 0.3 & syn & 0.4 & 0.19 \\
GdII & 4063.38 & 0.99 & 0.33 & syn & 0.4 & 0.18 \\
GdII & 4251.73 & 0.38 & -0.22 & syn & 0.35 & 0.07 \\
TbII & 3874.17 & 0.0 & 0.27 & syn & <-0.06 & -- \\
DyII & 3407.8 & 0.0 & 0.18 & syn & 0.3 & 0.09 \\
DyII & 3531.71 & 0.0 & 0.77 & syn & 0.32 & 0.19 \\
DyII & 3757.37 & 0.1 & -0.17 & syn & 0.37 & 0.06 \\
DyII & 3944.68 & 0.0 & 0.11 & syn & 0.17 & 0.04 \\
DyII & 4077.97 & 0.1 & -0.04 & syn & 0.15 & 0.05 \\
HoII & 3456.01 & 0.0 & 0.76 & syn & -0.42 & 0.07 \\
HoII & 4045.45 & 0.0 & -0.05 & syn & -0.3 & 0.07 \\
ErII & 3499.1 & 0.06 & 0.29 & syn & 0.22 & 0.07 \\
ErII & 3906.31 & 0.0 & 0.12 & syn & 0.01 & 0.12 \\
TmII & 3848.02 & 0.0 & -0.14 & syn & -0.67 & 0.06 \\
YbII & 3694.19 & 0.0 & -0.3 & syn & -0.27 & 0.05 \\
LuII & 2615.41 & 0.0 & 0.11 & syn & -0.7 & 0.17 \\
HfII & 3399.79 & 0.0 & -0.57 & syn & <0.67 & -- \\
OsII & 2282.28 & 0.0 & -0.57 & syn & 0.49 & 0.29 \\
IrI & 2639.71 & 0.0 & -0.31 & syn & <1.01 & -- \\
PtI & 2659.45 & 0.0 & -0.03 & syn & 0.53 & 0.1 \\
PtI & 2997.96 & 0.10 & -0.50 & syn & 0.68 & 0.16 \\
AuI & 2427.95 & 0.0 & -0.15 & syn & -0.1 & 0.12 \\
AuI & 2675.95 & 0.0 & -0.45 & syn & -0.02 & 0.11 \\
ThII & 4019.13 & 0.0 & -0.23 & syn & <-0.38 & -- \\
UII & 3859.57 & 0.04 & -0.07 & syn & <0.43 & -- \\
C-H & 4313.0 & -- & -- & syn & 6.7 & 0.12 \\
C-N & 3875.0 & -- & -- & syn & <7.73 & -- \\
\caption{\label{tab:all_lines} Wavelength, excitation potential, and oscillator strength of all absorption lines used are listed. Also listed is the abundance determination technique used i.e., spectral synthesis (syn) or EW measurement (ew). In the final columns, the abundance of the line and the associated systematic uncertainty are listed.}
\end{longtable}